\title{\LARGE A Bayesian hierarchical model for bridging across patient subgroups in phase I clinical trials with animal data}

\author{HAIYAN ZHENG$^{1, 3, \ast}$, LISA V. HAMPSON$^{2}$, THOMAS JAKI$^{3}$ \\[8pt]
\textit{$^1$Biostatistics Research Group, Population Health Sciences Institute, Newcastle University, U.K.} \\
\textit{$^2$Advanced Methodology and Data Science, Novartis Pharma AG, Switzerland.}\\
\textit{$^3$Department of Mathematics and Statistics, Lancaster University, U.K.}
\\[8pt]
$^\ast$Email: \href{mailto:haiyan.zheng@newcastle.ac.uk}{haiyan.zheng@newcastle.ac.uk}
}

\date{ }

\documentclass[11pt]{article}

\usepackage[margin=0.78in]{geometry}
\usepackage{hyperref}
\usepackage{amssymb, amsmath}
\usepackage{graphicx, caption}
\usepackage{booktabs}
\usepackage{enumerate}

\usepackage{cite}
\usepackage{natbib}
\usepackage{pdfpages}

\begin{document}
\maketitle

\begin{abstract}
Incorporating preclinical animal data, which can be regarded as a special kind of historical data, into phase I clinical trials can improve decision making when very little about human toxicity is known.
In this paper, we develop a robust hierarchical modelling approach to leverage animal data into new phase I clinical trials, where we bridge across non-overlapping, potentially heterogeneous patient subgroups.
Translation parameters are used to bring both historical and contemporary data onto a common dosing scale. This leads to feasible exchangeability assumptions that the parameter vectors, which underpin the dose-toxicity relationship per study, are assumed to be drawn from a common distribution. 
Moreover, human dose-toxicity parameter vectors are assumed to be exchangeable either with the standardised, animal study-specific parameter vectors, or between themselves.
Possibility of non-exchangeability for each parameter vector is considered to avoid inferences for extreme subgroups being overly influenced by the other.
We illustrate the proposed approach with several trial data examples, and evaluate the operating characteristics of our model compared with several alternatives in a simulation study.
Numerical results show that our approach yields robust inferences in circumstances, where data from multiple sources are inconsistent and/or the bridging assumptions are incorrect.
\end{abstract}


\section{Introduction}

Bridging strategies are increasingly being used in the paradigm of global drug development \citep{JBS:Huang2012, JBS:Tsong2012, JBS:Li2012, BMJ:Viergever2015}, aiming to minimise duplication of clinical research without disregarding heterogeneity between patient groups. Bridging studies may be conducted in a new geographic region such as Japan to evaluate similarity of the performance, typically with respect to efficacy, of a medicine which has likely been approved in other parts of the world, for instance, Europe, based on a complete clinical drug development program. The International Conference on Harmonisation \citet{EMA:ICH1998, EMA:ICH2006} discussed whether and when trial data generated in an `original' region can be leveraged to support the evaluation of drug activities in a new region, where a sponsor is seeking registration. 
The degree of borrowing, ranging from none to full, is a matter of negotiation between the sponsor and the local health authorities.
Bridging strategies can mitigate the drug lag problem \citep{DICP:Haen1975, PCR:Wileman2010, CPT:Ueno2013}, and expedite patient access to new medicines.  \\

Over the past few decades, the Pharmaceuticals and Medical Devices Agency (PMDA) in Japan promoted  synchronisation of clinical drug development in Japan and other countries \citep{JPMA:Principles2007}. The agency encourages domestic sponsors to participate in global phase I dose-finding studies, which has led to a number of bridging studies conducted in the early phase; see, for example, \citet{TIRS:Ogura2014}.
In this paper, we will focus on the design and analysis of phase I bridging studies, which aim to support estimation of the maximum tolerated dose (MTD) in a new geographic region or patient subgroup. We would thus like to leverage dose-toxicity data available in other but relevant studied populations.
In this setting, both {\em intrinsic} factors such as a patient's genetic make-up, and {\em extrinsic} factors such as diagnostic criteria as well as environmental exposures, can result in a different MTD across the regions or patient subgroups.
However, a review of 54 phase I oncology trials, conducted at the National Cancer Center Hospital in Japan between 1995 and 2012 \citep{JCO:Mizugaki2015}, found evidence of a small between-region heterogeneity in the toxicity profile of single cytotoxic agents. \\

Several model-based designs have been proposed for phase I clinical trials to address concerns about differences between patient subgroups.
\citet{SiM:Liu2015} develop a bridging continual reassessment method (CRM) procedure that uses the dose-toxicity data from a completed historical trial to generate multiple sets of `skeleton' probabilities for a new trial in another geographic region. The authors then use  the Bayesian model averaging \citep{JASA:Yin2009} to reconcile this information.
\citet{PS:Takeda2018} present a Bayesian dose-escalation procedure which dynamically leverages information from a historical study. Before the new trial begins, historical trial data are used to formulate a weakly informative prior for the parameter of a functional dose-toxicity model employed by the CRM for dose recommendations; so-called weakly informative because the prior effective sample size \citep{BIOM:Morita2008} is considerably smaller than the anticipated sample size of the new trial. Historical and new trial data are then linked through a `historical-to-current' parameter, which reflects the degree of agreement between studies.  \\

Alternatively, relevant `complementary-data' (or co-data for short) \citep{SIBR:Neuenschwander2016} can be drawn from other phase I clinical trials run concurrently to the trial of interest, or one could leverage data on a related patient subgroup enrolled in the same trial.
\citet{JBS:Quigley1999} propose a two-sample CRM to draw inferences about the MTDs for two non-overlapping groups of patients simultaneously. 
A shift model has been further discussed in the context of bridging studies by \citet{SBR:Quigley2014}, which constrains the recommended dose in the second subgroup to be identical to, or several levels shifted away the estimate in the first subgroup. \citet{PS:Wages2015} extended this shift model to design a phase I/II trial of stereotactic body radiation therapy, where uncertainty surrounding the `true shift' is tacitly concerned.
These authors have restricted attention to the scenario where co-data are exclusively the human toxicity data, collected from other subgroups. However, no phase I clinical trial is planned in a vacuum: preliminary data characterising the toxicity profile of the drug will typically be available in animals, as is required by regulatory authorities \citep{FDA:FIH2005}. It thus appears to be appealing to use this information, such that dose recommendations at early stages of the trials can be backed up with sufficient evidence \citep{SMMR:Zheng2019, BJ:Zheng2019}.  \\

\citet{SMMR:Zheng2019} propose a Bayesian hierarchical model to leverage data from multiple animal species in a phase I oncology trial, which will be performed in a homogeneous patient group, to support the interim and final dosing recommendations. In this paper, we extend their approach to accommodate the circumstances where the study population may involve heterogeneous patient subgroups. Specifically, the robust extention proposed in this paper can augment a phase I bridging trial with co-data, which may comprise (i) data from completed preclinical animal studies, (ii) concurrent external data from either completed or ongoing trials conducted in related patient subgroups (e.g., patients from other geographic regions). When the intrinsic and extrinsic factors arising from ethnicity would modify the dose-toxicity relationship, our model will estimate the MTD specific to the regions. \\



The remainder of this paper is structured as follows. In Section \ref{sec:robBHM}, we develop a robust Bayesian hierarchical model to leverage animal data into phase I trials, which can address potential between-subgroup heterogeneity. In Section \ref{sec:illex}, we show several illustrative examples using our methodology to improve decision making in phase I trials with co-data. In Section \ref{sec:sims}, we perform a simulation study to evaluate the operating characteristics of dose-escalation trials following the proposed design in comparison with several alternatives. Finally, we draw conclusions and look towards future research in Section \ref{sec:discuss}.

\section{Bayesian hierarchical modelling for data from heterogeneous sources}
\label{sec:robBHM}

In this section, we generalise the Bayesian model of \citet{SMMR:Zheng2019} to leverage available animal data and dose-toxicity data from different human subgroups into new phase I clinical studies. \\

Suppose that at the time of planning a phase I clinical trial, $M$ preclinical studies have been performed in $K$ animal species, labelled $S_1, \dots, S_K$.
For $i = 1, \dots, M$, animal study $i$ tested a total of $J_i$ doses contained in set $\mathcal{D}_i = \{d_{i1}, \dots, d_{iJ_i}; d_{it_1} \leq d_{it_2} \text{ for } 1 \leq t_1 \leq t_2 \leq J_i \}$.
On receiving dose $d_{ij} \in \mathcal{D}_i$, an animal experiences a dose-limiting toxicity (DLT) with probability $p_{ij}$ and no DLT with probability $1-p_{ij}$.
Let $n_{ij}$ and $r_{ij}$ be the number of animals that received dose $d_{ij}$ and the number that experienced a DLT, respectively.
We assume a monotonic increasing relationship between $p_{ij}$ and $d_{ij}$, which can be adequately described by a two-parameter logistic regression model \citep{BioPS:Whitehead1998, SIM:Neuenschwander2008}:
\begin{equation}
\begin{split}
r_{ij} | p_{ij}, n_{ij} &\sim \text{Binomial}(p_{ij}, n_{ij}),  \text{  for } j = 1, \dots, J_i, \\
\text{logit}(p_{ij}) &= \theta_{1i} + \exp(\theta_{2i}) \log(\delta_{\mathcal{A}_i} d_{ij} / d_\text{Ref}),
\end{split}
\label{eq:dose-tox}
\end{equation} 
\noindent where $\delta_{\mathcal{A}_i}$ is a translation parameter mapping animal doses onto an equivalent human dosing scale.
\citet{SMMR:Zheng2019} propose placing a tailored log-normal prior on $\delta_{\mathcal{A}_i}$ to account for the intrinsic differences between the toxicity profile of the drug in animal species $\mathcal{A}_i \in \{S_1, \dots, S_K\}$ and humans. 
Thus, model parameters $\boldsymbol{\theta}_i = (\theta_{1i}, \theta_{2i})$ describe the dose-toxicity relationship  (equivalently) in humans. 
In Model \eqref{eq:dose-tox}, $d_\text{Ref}$ is a reference dose invariant across all dose-toxicity studies, which is often chosen to be the likely human MTD. \\

Random-effects distributions are stipulated on the second level of the hierarchical model to enable information sharing between animal studies of the same species: 
\begin{equation}
\boldsymbol{\theta}_i | \boldsymbol{\mu}_{\mathcal{A}_i}, \Psi \sim \text{BVN}(\boldsymbol{\mu}_{\mathcal{A}_i}, \Psi),
\label{eq:parmod}
\end{equation}
\noindent with
\begin{align*}
\begin{split}
\boldsymbol{\mu}_{\mathcal{A}_i} = \begin{pmatrix} 
\mu_{1 \mathcal{A}_i}  \\
\mu_{2 \mathcal{A}_i}
\end{pmatrix}
\quad {\rm and} \quad
\Psi = \begin{pmatrix} 
\tau_{1}^2 & \rho \tau_{1} \tau_{2}\\
\rho \tau_{1} \tau_{2} & \tau_{2}^2
\end{pmatrix}
\end{split}
\end{align*}
\noindent for $\mathcal{A}_i \in \{S_1, \dots, S_K \}$. 
Variances in $\Psi$ reflect the magnitude of between-study heterogeneity within an animal species.
A `supra-species' random effects distribution is introduced to facilitate borrowing of information across different animal species. That is, for species $S_k, \, k = 1, \dots, K$,
\begin{equation}
\boldsymbol{\mu}_{S_k} | \boldsymbol{m}, \Sigma \sim \text{BVN}(\boldsymbol{m}, \Sigma), 
\label{eq:suprapar}
\end{equation}

\noindent with

\begin{align*}
\begin{split}
\boldsymbol{m} = \begin{pmatrix} 
m_1  \\
m_2
\end{pmatrix}
\quad {\rm and} \quad
\Sigma = \begin{pmatrix} 
\sigma_{1}^2 & \kappa \sigma_{1} \sigma_{2}\\
\kappa \sigma_{1} \sigma_{2} & \sigma_{2}^2
\end{pmatrix}. 
\end{split}
\end{align*}

\noindent This `supra-species' random-effects distribution accounts for the differences between toxicity parameters in different species which are not addressed by the translation parameters $\delta_{S_1}, \dots, \delta_{S_K}$.  \\

We now focus on modelling the human toxicity data that will be collected from different human subgroups. 
Suppose there are a total of $L$ predefined, non-overlapping human subgroups and one trial only is performed in each subgroup.
To distinguish from the notation used for animal studies, we let $\ell=1, \dots, L$ index the new human trials wherein the doses $d_{\ell j} \in \mathcal{D}_{\ell} = \{d_{\ell 1}, \dots, d_{\ell J_\ell}; d_{\ell t_1} \leq d_{\ell t_2} \text{ for } 1 \leq t_1 \leq t_2 \leq J_\ell \}$ are to be evaluated, and let $\boldsymbol{\gamma}_\ell = (\gamma_{1\ell}, \gamma_{2\ell})$ be the counterpart of $\boldsymbol{\theta}_i$; that is, $\boldsymbol{\gamma}_\ell$ underpins the dose-toxicity relationship in human subgroup $\ell$.
Model \eqref{eq:dose-tox} is also applicable to describe the human toxicity data, only that we will set the animal-to-human translation parameter $\delta_{\mathcal{A}_i}=1$ and introduce a subgroup-specific parameter denoted by $\epsilon_{\ell}$ for subgroup $\ell$.
For a phase I clinical trial $\ell = 1, \dots, L$, the human toxicity data can be described by
\begin{equation}
\begin{split}
r_{\ell j} | p_{\ell j}, n_{\ell j} &\sim \text{Binomial}(p_{\ell j}, n_{\ell j}),  \text{  for } j = 1, \dots, J_{\ell}, \\
\text{logit}(p_{\ell j}) &= \gamma_{1\ell } + \exp(\gamma_{2\ell }) \log(\epsilon_{\ell} d_{\ell j} / d_\text{Ref}),
\end{split}
\label{eq:Bdose-tox}
\end{equation}
\noindent where $d_\text{Ref}$ is the same reference dose used in Model (\ref{eq:dose-tox}) and $\epsilon_{\ell}$ adjusts for the differences in toxicity arising from the intrinsic and/or extrinsic factors across human subgroups.
In particular, this parameterisation maps the dose-toxicity profiles of different subgroups onto an `average' human dosing scale.
We regard each $\epsilon_\ell$ as a random variable, on which we place a normal prior distribution $N(1, \nu^2_\ell)$, further truncated for positive real numbers only, due to the use of a logrithm. \\

Here, we use a truncated normal prior symmetric about the value 1 instead of a log-normal prior on $\epsilon_\ell$ as we do for the animal-to-human translation parameter $\delta_{S_k}$'s. Because the dose-toxicity data collect from each subgroup $\ell$ can appropriately be assumed on the `average' human dosing scale. More importantly, it gives equal probability mass on $\epsilon_\ell >1$ and $0 < \epsilon_\ell < 1$, which corresponds to the drug at the same dose being more or less toxic than the average in subgroup $\ell$, respectively. To strictly ensure the symmetry, the normal prior will be truncated for values bounded by 0 and $U_\nu$, which depends on the choice of $\nu_\ell$. 
An example is to place a normal prior $N(1, 0.255^2)$ truncated to fall within (0, 2) on each $\epsilon_\ell$, such that the 95\% probability mass is concentrated on the interval [0.5, 1.5]. 
This indicates the region-specific MTDs, if divergent, are expected to have less than 0.5-fold change between one another.
Our stipulation here is consistent with the general consensus that the toxicity risk would rarely be more than doubled in another ethnic subgroup if the bridging assumption is correct. \\

Following \citet{SIBR:Neuenschwander2016} who allow for the possibility of more than one exchangeability distribution, our model accommodates two exchangeability scenarios for $\boldsymbol{\gamma}_1, \dots, \boldsymbol{\gamma}_L$. That is, a) parameters of animal and human dose-toxicity relationships are exchangeable with each other; and b) human dose-toxicity parameters are exchangeable only with those of other human subgroups.
For human subgroup $\ell = 1, \dots, L$, we stipulate that

\begin{enumerate}[(i)]
\item  with prior probability $w_{\ell S_k}$ for $k = 1, \dots, K$:
$$
\boldsymbol{\gamma}_{\ell} | \boldsymbol{\mu}_{S_k}, \Psi \sim {\rm BVN}(\boldsymbol{\mu}_{S_k}, \Psi);
$$
This represents exchangeability between $\boldsymbol{\gamma}_\ell$ and the study-specific parameters $\boldsymbol{\theta}_i$ relating to animal species $S_k$.
\item  with prior probability $w_{\ell \mathcal{H}}$:
\begin{equation}
\boldsymbol{\gamma}_{\ell} | \boldsymbol{\mu}_\mathcal{H}, \Phi \sim {\rm BVN}(\boldsymbol{\mu}_\mathcal{H}, \Phi), 
\label{eq:bridgingpar}
\end{equation}

\noindent so that $\boldsymbol{\gamma}_\ell$ is exchangeable only with the parameter vectors of the other human subgroups and where

\begin{align*}
\begin{split}
\boldsymbol{\mu}_\mathcal{H} = \begin{pmatrix} 
\mu_{1\mathcal{H}}  \\
\mu_{2\mathcal{H}}
\end{pmatrix}
\quad {\rm and} \quad
\Phi = \begin{pmatrix} 
\tau_{3}^2 & \eta \tau_{3} \tau_{4}\\
\eta \tau_{3} \tau_{4} & \tau_{4}^2
\end{pmatrix};
\end{split}
\end{align*}

\item  with prior probability $w_{\ell R} = 1 - \sum_k w_{\ell S_k} - w_{\ell \mathcal{H}}$: 
$$
\boldsymbol{\gamma}_{\ell}  \sim {\rm BVN}(\boldsymbol{m}_{0\ell}, R_{0\ell}), 
$$
so that $\boldsymbol{\gamma}_\ell$ is non-exchangeable with any other dose-toxicity parameters. 
\end{enumerate}

\noindent
Before the conduct of a phase I trial in subgroup $\ell$, we need to pre-specify the prior probabilities $w_{\ell S_1}, \dots, w_{\ell S_K}$, $w_{\ell \mathcal{H}}$ and $w_{\ell R}$.
The judgements of translational scientists or pharmacologists will be valuable. 
Stipulating a large $w_{\ell S_k}$ or $w_{\ell \mathcal{H}}$ reflects a high level of prior confidence in the relevance of data from animal species $S_k$ or in the bridging assumption.  \\

This second `human only' exchangeability distribution has its own covariance matrix $\Phi$. This is because the degree of heterogeneity between study-specific dose-toxicity parameters in humans may be quite different to the level of heterogeneity between study-specific parameters in animals or the variations across species, captured by $\Psi$ and $\Sigma$, respectively.
When animal data have very limited predictability of the human toxicity yet the human toxicity data between themselves share considerable commonality, our robust hierarchical model will lead to large posterior probabilities attributed to the $(K+1)$th `human only' exchangeability distribution. 
For additional robustness, the model assigns positive prior probabilities $w_{\ell R}$ to the case that each $\boldsymbol{\gamma}_\ell$ is not exchangeable with any other parameter vectors. 
When the dose-toxicity relationship of a human subgroup appears to be an outlier, that is, is dissimilar to that of any other human subgroups or animal species, the dose-toxicity parameters can be estimated based on their own independent prior ${\rm BVN}(\boldsymbol{m}_{0\ell}, R_{0\ell})$. \\

To complete our Bayesian model, we now specify priors for other parameters.
Weakly informative priors are placed on the hyperparameters of the random effects distributions in Model \ref{eq:suprapar} and Model \ref{eq:bridgingpar}.
The weakly informative priors used in subsequent sections are chosen so that each human toxicity risk $p_{\ell j}$ has a wide 95\% prior credible interval \citep{AAS:Gelman2008}.
For the `supra-species' population means $\boldsymbol{m} = (m_1, m_2)$, we set $m_1 \sim N(b_1, s_1^2)$ and $m_2 \sim N(b_2, s_2^2)$. 
The same normal priors are used for $\mu_{1\mathcal{H}}$ and $\mu_{2\mathcal{H}}$, respectively.
Priors for the variance parameters should reflect opinion on the degree of between-source heterogeneity.
Here, we propose setting
\begin{equation}
\label{eq:hyperVar}
\begin{split}
\tau_1 &\sim HN(z_1),   \quad \tau_2  \sim HN(z_2),   \quad \tau_3 \sim HN(z_3),   \quad \tau_4  \sim HN(z_4),     \\
\sigma_1 &\sim HN(c_1),   \quad \sigma_2  \sim HN(c_2),   \quad \rho \sim U(-1, 1), \quad \kappa \sim U(-1, 1), \quad  \eta \sim U(-1, 1),
\end{split}
\end{equation}

\noindent where $HN(z)$ denotes a half-normal distribution formed by truncating a normal distribution $N(0, z^2)$ to fall within ($0, \infty$). 
This robust Bayesian hierarchical model can be fitted using Markov chain Monte Carlo. We provide the OpenBUGS \citep{SiM:Lunn2009} code in Appendix \ref{sec:appdCode} for the implementation of our Bayesian analysis model.

\section{Illustrative example}
\label{sec:illex}

In this section, we apply the robust Bayesian hierarchical model proposed in Section \ref{sec:robBHM} to a hypothetical trial example based on a real trial, which aimed to characterise the toxicity profile of GSK3050002 \citep{GSK:NCT02671188}, an antibody for treating patients with psoriatic arthritis. 
The original trial enrolled a total of 49 human subjects exclusively recruited from the United Kingdom, but we assume that two hypothetical phase I trials (labelled $\mathcal{T}_1$ and $\mathcal{T}_2$) are to be performed in two geographic regions, $\mathcal{R}_1$ and $\mathcal{R}_2$, respectively. 
For illustrative purposes, we suppose the trial $\mathcal{T}_2$ will be performed in region $\mathcal{R}_2$ after the trial data of $\mathcal{T}_1$ in $\mathcal{R}_1$ are made available. The co-data for trial $\mathcal{T}_2$ thus comprise the data of trial $\mathcal{T}_1$ and animal studies where possible. The choice of animal species, animal doses and human doses for our numerical studies are informed by the real GSK phase I clinical trial. 
For present purposes, the principal aim of these hypothetical trials is to estimate a region-specific MTD, defined as the dose associated with a risk of DLT of 25\%.

\subsection{Hypothetical preclinical data and predictive priors for human toxicity}
\label{sec:hypoAni}

According to the protocol of GSK3050002 \citep{GSK:PhIprotocol}, preclinical toxicity studies have been performed in monkeys and rats.
Moreover, monkeys were thought to be the most relevant animal species for predicting toxicity in humans of this drug. 
In the two real monkey studies, doses 1, 10, 30, 100 mg/kg were tested on 4 -- 12 monkeys per dose group.
From the trial protocol, it was not possible to identify what dose levels were used in rats, nor the exact number of rats treated, nor the number of toxicities observed. 
We therefore simulate plausible animal datasets based on the limited information available, and use these simulated data to obtain predictive priors for the human toxicity at doses contained in set $\mathcal{D}_{\ell} = \{0.1, 0.5, 1, 5, 10, 20 \}$ mg/kg, which will be evaluated in trials $\mathcal{T}_1$ and $\mathcal{T}_2$.
The simulated animal data are represented in Figure S1 of the Web-based Supplementary Materials. \\

Throughout, we set $d_{\rm Ref} = 5$ mg/kg and use the priors as follows.
We consider $m_1 \sim N(-1.099, 1.98^2)$ and $m_2 \sim N(0, 0.99^2)$ for the `supra-species' population means as well as for the elements of $\boldsymbol{\mu}_\mathcal{H}$, and $\sigma_1 \sim HN(1)$ and $\sigma_2 \sim HN(0.5)$ for the variances that permit sharing of information across animal species. 
We assume moderate-to-substantial between-study heterogeneity between animal studies of the same species would make sense, and therefore let $\tau_1 \sim HN(0.5), \tau_2 \sim HN(0.25)$.
Further, we assume small-to-moderate heterogeneity between ethnic subgroups, captured by $\tau_3 \sim HN(0.25), \tau_4 \sim HN(0.125)$.  
Here, we stipulate a half-normal prior $HN(z)$ with smaller $z$ for the slope than that for the intercept, because we desire to borrow more in terms of the shape rather than the location of the dose-toxicity curves \citep{Kamrin1988}.
Following \citet{SMMR:Zheng2019}, we set $\delta_{\rm Rat} \sim LN(-1.820, 0.323^2)$ and $\delta_{\rm Monkey} \sim LN(-1.127, 0.273^2)$ to translate the animal data onto a common human scale.  \\

\begin{figure}[!htb]
\centering
\captionsetup{font=scriptsize}
\includegraphics[width=0.8\linewidth]{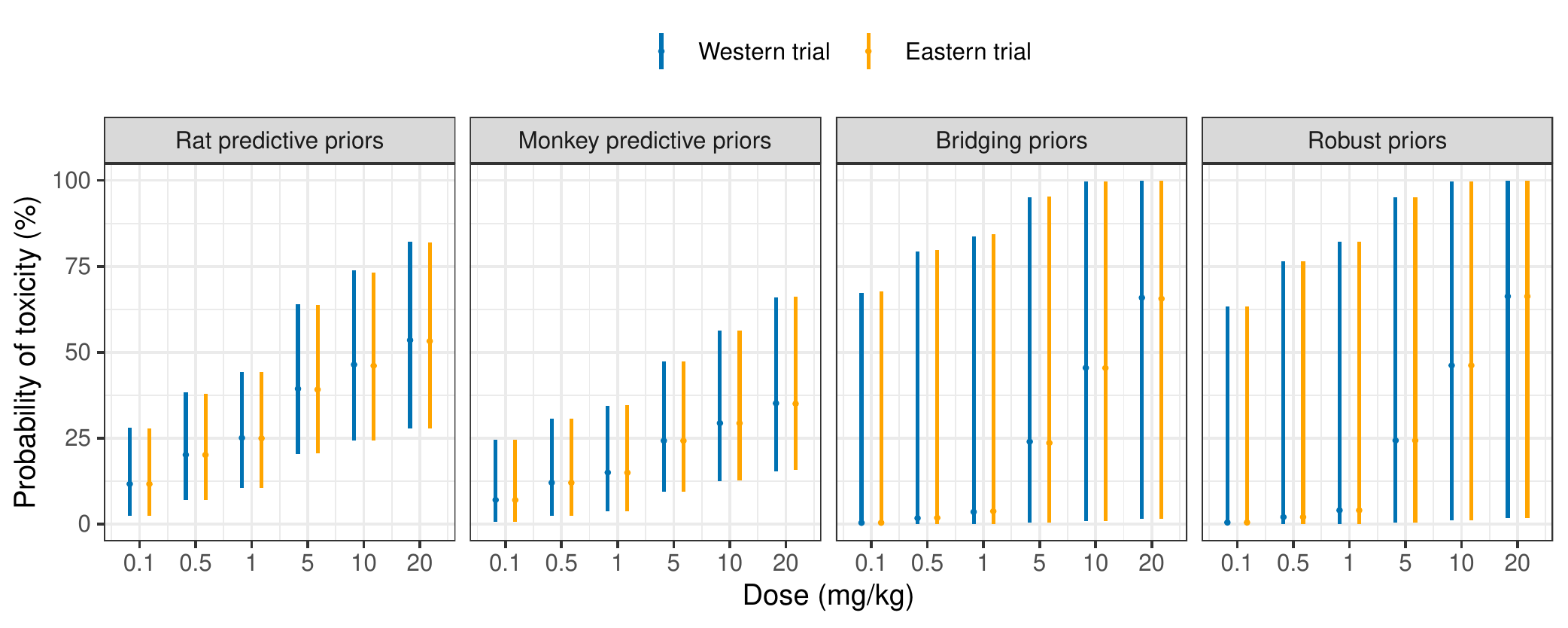}
\caption{Summaries about the predictive priors for human toxicity, when using animal data from a single species (Panels A and B) or no animal data at all (Panel C). Medians together with 95\% credible intervals of the marginal predictive priors are plotted.}
\label{fig:spPriors}
\end{figure}

For a robust inference under scenarios of data inconsistency, the non-exchangeability distributions BVN$(\boldsymbol{m}_{0\ell}, R_{0\ell})$ are specified for each trial $\ell$, independently.
To be more specific, we set $m_{01\ell} \sim N(-1.099, 2^2)$ and $m_{02\ell} \sim N(0, 1^2)$, with a zero correlation between $m_{01\ell}$ and $m_{02\ell}$. 
By specifying the prior probability of exchangeability as 1 to a specific animal species and retaining the rest as 0, predictive priors on toxicity $p_{\ell j}$ for each phase I trial can be obtained based on animal data of a single species. 
For example, to see how human toxicity may be predicted by the monkey datasets, we may fix $w_{\ell {\rm Rat}}=0,\, w_{\ell {\rm Monkey}}=1,\, w_{\ell \mathcal{H}}=0, w_{\ell R}=0$. 
When we set $w_{\ell R}=1$ and retain the rest as 0, no animal data will be used, nor are we making an assumption of bridging.
Figure \ref{fig:spPriors} presents the key summaries of such predictive priors by the source of information. 
This would be useful to examine whether our Bayesian model can borrow (discount) information quickly from a particular species, given the data consistency (inconsistency). \\

[Figure \ref{fig:spPriors} about here.] \\

As we can see, the rat and monkey data predict 1 mg/kg and 5 mg/kg as doses highly likely to result in a DLT risk close to 25\% in a human trial.
After translation of the animal doses, rat data are mainly projected on the low doses of $\mathcal{D}_{\ell}$. Predictive priors obtained solely from rat data are thus more diffuse at high doses such as 10 mg/kg and 20 mg/kg, at which monkey data in contrast have produced predictive priors for the DLT risks with narrower credible intervals. 
Patients recruited in regions $\mathcal{R}_1$ and $\mathcal{R}_2$ are predicted as having similar DLT risks based on the animal data.
This is because the same prior probabilities $w_{\ell S_k}, \, w_{\ell \mathcal{H}}$ and $w_{\ell R}$, as well as the same truncated normal prior on $\epsilon_\ell$,  have been chosen for each human trial $\ell$ at the outset. \\

We obtain marginal predictive priors for the DLT risk in humans,by allocating prior weights to different animal species on the basis of their {\em a priori} predictability of the human toxicity.
For trial $\mathcal{T}_1$, we stipulate $w_{1{\rm Rat}} = 0.2, \, w_{1{\rm Monkey}} = 0.6,\, w_{1\mathcal{H}} = 0$ and $w_{1R} = 0.2$. 
No prior probability has been allocated to the exchangeability distribution for bridging across patient subgroups, because trial $\mathcal{T}_2$ has not yet started, and the co-data for $\mathcal{T}_1$ are exclusively from animal studies. We note this is the Bayesian model proposed by \citet{SMMR:Zheng2019}, suitably for leveraging animal data to one homogeneous patient group.
Figure S2 of the Supplementary Materials gives summaries about the marginal predictive priors that synthesise information across animal species. As soon as the trial $\mathcal{T}_2$ begins, we let the $(K+1)$th exchangeability component come into play.
For illustration, we set $w_{2{\rm Rat}} = 0.1, \, w_{2{\rm Monkey}} = 0.5,\, w_{2\mathcal{H}} = 0.2$ and $w_{2R} = 0.2$ to leverage both animal data and the $\mathcal{T}_1$ trial data, the human data from region $\mathcal{R}_1$, into trial $\mathcal{T}_2$. \\

We consider characterising each dose $d_{\ell j} \in \mathcal{D}_\ell$ with three interval probabilities; specifically, a dose may lead to patients being 
(i) underdosed, if the DLT risk is less than 0.16,
(ii) properly dosed, if the DLT risk fall within the target interval [0.16, 0.33), 
and (iii) overdosed, if the DLT risk is greater than 0.33 \citep{SIM:Neuenschwander2008}.
In this data example, we suggest choosing 0.1 mg/kg to be the safe starting dose for the first-in-man trial $T_1$, given $\mathbb{P}(p_{11}< 0.16 | \boldsymbol{Y}_1, \dots, \boldsymbol{Y}_5 ) = 0.872$.
Wherease, the safe starting dose for trial $\mathcal{R}_2$ will be determined based on both the animal data and the western human toxicity data with a similar approach. \\


It will be helpful to assess the effective sample size (ESS) \citep{BIOM:Morita2008} about the predictive priors before using them.
Before the conduct of human trials $\mathcal{T}_1$ and $\mathcal{T}_2$, we approximate each marginal predictive prior for DLT risk per dose by beta distributions with parameters $a$ and $b$, for the convenience of calculating the ESS as $(a+b)$. The parameters $a$ and $b$ are determined by matching the first two moments of a Beta$(a, b)$ with the original marginal predictive priors, obtained based on animal data for each $p_{\ell j}$.
Table S1 of the Web-based Supplementary Materials reports the computed ESSs. Basically, the animal data are equivalent to what would be acquired from 4.5 -- 8.2 human subjects treated in each trial. 
It may be worth to re-assess the ESSs after the completion of trial $\mathcal{T}_1$ to see how informative the co-data for trial $\mathcal{T}_2$ would be.

\subsection{Design and conduct of the phase I trials in different patient subgroups}

Suppose that the phase I trials $\mathcal{T}_1$ and $\mathcal{T}_2$ are planned to have equal sample size, say, 24 patients.
We begin by recruiting patients in cohorts of size three to trial $\mathcal{T}_1$.
After the toxicity responses have been observed from the last cohort of trial $\mathcal{T}_1$, the trial $\mathcal{T}_2$ begins with the same trial structure. 
Specifically, the co-data for trial $\mathcal{T}_2$ are from both the animal studies and trial $\mathcal{T}_1$.
We use $h_{\ell}$ to index the cohort number of trial $\mathcal{T}_\ell$, and 
$\boldsymbol{Y}_{\mathcal{T}_1}^{(h_{\ell})}$ or $\boldsymbol{Y}_{\mathcal{T}_2}^{(h_{\ell})}$ to denote the human toxicity data accrued up to the cohort $h_{\ell}$ of trial $\mathcal{T}_1$ or $\mathcal{T}_2$. \\

Recall that we have estimated dose 0.1 mg/kg as a suitable starting dose for patients in cohort $h_{\ell}=1$ of trial $\mathcal{T}_1$. For the subsequent patient cohorts $h_{\ell}\geq 2$ of the same trial, a dose will be recommended according to the criterion:
\begin{equation}
\hat{d}_{\mathcal{T}_1}^{(h_{\ell})} = \max \{ d_{\ell j} \in \mathcal{D}_{\ell}: \mathbb{P}(p_{\ell j} \geq 0.33 | \boldsymbol{Y}_1, \dots, \boldsymbol{Y}_5, \boldsymbol{Y}_{\mathcal{T}_1}^{(h_{\ell}-1)} ) \leq 0.25 \},
\label{eq:escalationW} 
\end{equation}

\noindent where $\boldsymbol{Y}_1, \dots, \boldsymbol{Y}_5$ represent the five hypothetical animal datasets collected from the rat and monkey studies. 
The phase I trials will be terminated either after completion of treatment for all the 24 patients, or for safety if the posterior risk of overdosing is too high.
When the complete data of trial $\mathcal{T}_1$, denoted by $\boldsymbol{Y}_{\mathcal{T}_1}$, are made available, we start the trial $\mathcal{T}_2$ with dose
$$
\hat{d}_{\mathcal{T}_2}^{(h_{\ell} = 1)} = \max \{ d_{\ell j} \in \mathcal{D}_{\ell}: \mathbb{P}(p_{\ell j} < 0.16 | \boldsymbol{Y}_1, \dots, \boldsymbol{Y}_5, \boldsymbol{Y}_{\mathcal{T}_1}) > 0.85 \}, 
$$
and a dose  
\begin{equation}
\hat{d}_{\mathcal{T}_2}^{(h_{\ell}\geq 2)} = \max \{ d_{\ell j} \in \mathcal{D}_{\ell}: \mathbb{P}(p_{\ell j} \geq 0.33 | \boldsymbol{Y}_1, \dots, \boldsymbol{Y}_5, \boldsymbol{Y}_{\mathcal{T}_1}, \boldsymbol{Y}_{\mathcal{T}_2}^{(h_{\ell}-1)} ) \leq 0.25 \}.
\label{eq:escalationE}
\end{equation}

\noindent to be recommended to patients in the subsequent cohorts.
To prevent escalating doses too rapidly, additional constraints such as ``never skipping a dose during the escalation'' may be applied in practice. This means, in our illustrative example, one cannot skip dose 0.5 mg/kg to recommend 1 mg/kg for patients in cohort $h_{\ell}=2$, even if the first three doses all comply with criteria \eqref{eq:escalationW} and \eqref{eq:escalationE}.  \\

\begin{figure}[!htb]
\centering
\captionsetup{font=scriptsize}
\includegraphics[width=0.9\linewidth]{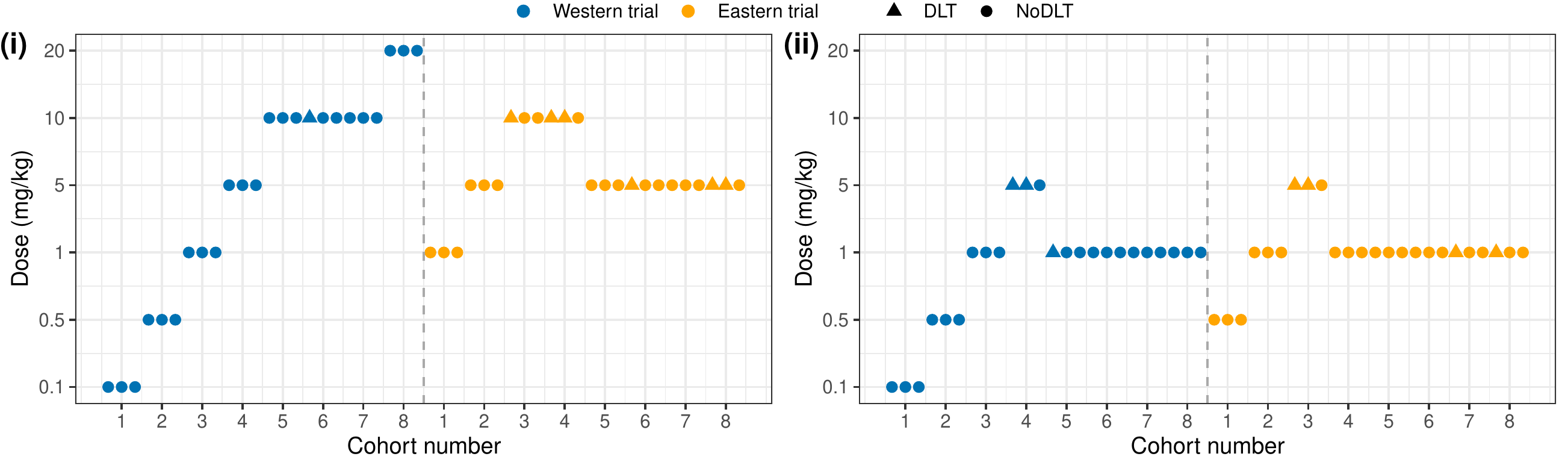}
\caption{Trial trajectory of hypothetical phase I trials performed in two geographic regions, in which trial data were simulated from (i) a
divergent scenario and (ii) a consistent scenario, respectively.}
\label{fig:DataSc}
\end{figure}

[Figure \ref{fig:DataSc} about here.] \\

Figure \ref{fig:DataSc} shows two simulated realisations of trials $\mathcal{T}_1$ and $\mathcal{T}_2$. 
These data examples were simulated under different scenarios for the human dose-toxicity relationship, basing dose recommendations on the proposed Bayesian model.
Specifically, subfigure (i) considers the scenario of divergent toxicity profiles in each human subgroup, while subfigure (ii) assumes the two relationships are consistent.
Prior specifications as well as the prior probabilities of exchangeability and non-exchangeability follows what was described in Section \ref{sec:hypoAni}.
Figure \ref{fig:DataSc} verifies that the choice of a safe dose to start with in trial $\mathcal{T}_2$ indeed relies on the toxicity data accrued from the first trial  $\mathcal{T}_1$. 
Reading subfigure (i) of Figure \ref{fig:DataSc} together with Figure \ref{fig:spPriors}, there seems to be no consistent animal data for the first-in-man trial $\mathcal{T}_1$ in scenario (i); moreover, considerable heterogeneity exists between trials $\mathcal{T}_1$ and $\mathcal{T}_2$.
Regardless of the sequential conduct of these phase I trials, our Bayesian approach allows any inconsistent information from external sources to be discounted quickly, leading to declaration of doses 20 mg/kg and 5 mg/kg as the region-specific MTDs. \\

On the completion of trial $\mathcal{T}_1$, we re-assess the ESSs before the start of trial $\mathcal{T}_2$. Table \ref{tab:pseudoptsT2} lists the corresponding ESSs for the marginal posteriors or priors for the human toxicity, given the hypothetical $\mathcal{T}_1$ trial data simulated from scenarios (i) and (ii), respectively. From this, even when there exists fairly rich information on human toxicity from a different patient subgroup, the predictive priors for trial $\mathcal{T}_2$ obtained using the proposed Bayesian model are very unlikely to dominate the analysis.

\section{Simulation study}
\label{sec:sims}

In this section, we compare the operating characteristics of phase I dose-escalation trials, conducted using the proposed Bayesian hierarchical model or an alternative Bayesian model. 
The Bayesian analysis models we consider are as follows.

\begin{itemize}
\item Model A is the proposed Bayesian model that leverages co-data from multiple sources;
\item Model B discards animal data; human parameter vectors $\boldsymbol{\gamma}_1$ and $\boldsymbol{\gamma}_2$ are assumed fully exchangeable;
\item Model C analyses trials $\mathcal{T}_1$ and $\mathcal{T}_2$ separately, without leveraging any animal data;
\item Model D leverages animal data to trials $\mathcal{T}_1$ and $\mathcal{T}_2$, with $w_{\ell {\rm Rat}} = 0.2, \, w_{\ell {\rm Monkey}} = 0.6$ and $w_{\ell R} = 0.2$, respectively, but permits no borrowing across human subgroups;
\item Model E analyses trial $\mathcal{T}_1$ without using any co-data, and trial $\mathcal{T}_2$ pooling data only from $\mathcal{T}_1$.
\end{itemize}

The prior specifications for Model A remain unchanged from Section \ref{sec:hypoAni}.
All the simulated $\mathcal{T}_1$ trials, regardless of the analysis model, begin with the lowest dose 0.1 mg/kg.
Simulated $\mathcal{T}_2$ trials begin with dose 0.1 mg/kg, when implementing Models C and D.
However, the choice of a safe starting dose for $\mathcal{T}_2$ is conditional on both animal data and data from $\mathcal{T}_1$ trial data when using analysis Model A, and solely on $\mathcal{T}_1$ trial data when using Model B or E.
In these settings, we select as the starting dose for trial $\mathcal{T}_2$ the highest dose $d_{\ell j^\star}$ that complies with $\mathbb{P}(p_{\ell j^\star} < 0.16 | \boldsymbol{Y}_1, \dots, \boldsymbol{Y}_5, \boldsymbol{Y}_{\mathcal{T}_1}) > 0.85$ for Model A, or $\mathbb{P}(p_{\ell j^\star} < 0.16 | \boldsymbol{Y}_{\mathcal{T}_1}) > 0.85$ for Model B or E.
Model D is essentially the Bayesian model of \citet{SMMR:Zheng2019} for incorporating animal data into a phase I trial in a homogeneous human population.
We note that Model A simplifies to Model D if $w_{\ell \mathcal{H}} = 0$. \\

[Table \ref{tab:pToxH} about here.] \\

Each simulated phase I trial is performed in an adaptive manner: interim dose recommendations are made according to criteria \eqref{eq:escalationW} and \eqref{eq:escalationE} for trials $\mathcal{T}_1$ and $\mathcal{T}_2$, respectively.
Given the true probability of toxicity per human dose listed in Table \ref{tab:pToxH}, we simulate the DLT outcomes of patients from a binary distribution.
We evaluate the properties of the Bayesian models under six scenarios, comprising cases where there are conflicts across data sources, and cases where parameters in different subgroups are exchangeable.
Scenarios 1 and 6 represent two extremes. 
Only simulated trials with all 24 patients treated and their toxicity outcomes observed will lead to a declaration of a region-specific MTD. At the end of a complete trial in region $R_\ell$, we declare as the MTD the dose
$$
\hat{d}_{\ell{\rm M}} = \arg \underset{d_{\ell j}\in \mathcal{D}_\ell'}{\min} |\tilde{p}_{\ell j} - 0.25|,
$$
where $\tilde{p}_{\ell j}$ denotes the posterior median DLT risk at dose $d_{\ell j}$, and $\mathcal{D}_{\ell}'\subseteq \mathcal{D}_\ell$ contains all the doses that were used to treat patients in the study, and also satisfy our overdose criterion.
Simulations were run in R (version 3.4.4) \citep{RSoftware} using the package R2OpenBUGS \citep{R2OpenBUGS} based on two parallel chains, each contributing  15 000 MCMC samples and sacrificing the first 5 000 iterations as burn-in. 
\\

For each toxicity scenario, we simulated 1000 pairs of adaptive phase I dose-escalation trials in regions $\mathcal{R}_1$ and $\mathcal{R}_2$ per Bayesian model.
Results in the following were summarised by region.
Averaging across the simulated phase I trials, we reported the percentage of trials that were stopped early for safety, and percentage of trials declaring a specific dose as MTD. In addition, we reported the average number of patients that were allocated to each dose. This helps us to understand whether different Bayesian models carry greater risks for patients' safety particularly in scenarios of excessive toxicity, such as scenario 5.   \\

\begin{figure}[!htb]
\centering
\captionsetup{font=scriptsize}
\includegraphics[width=0.85\linewidth]{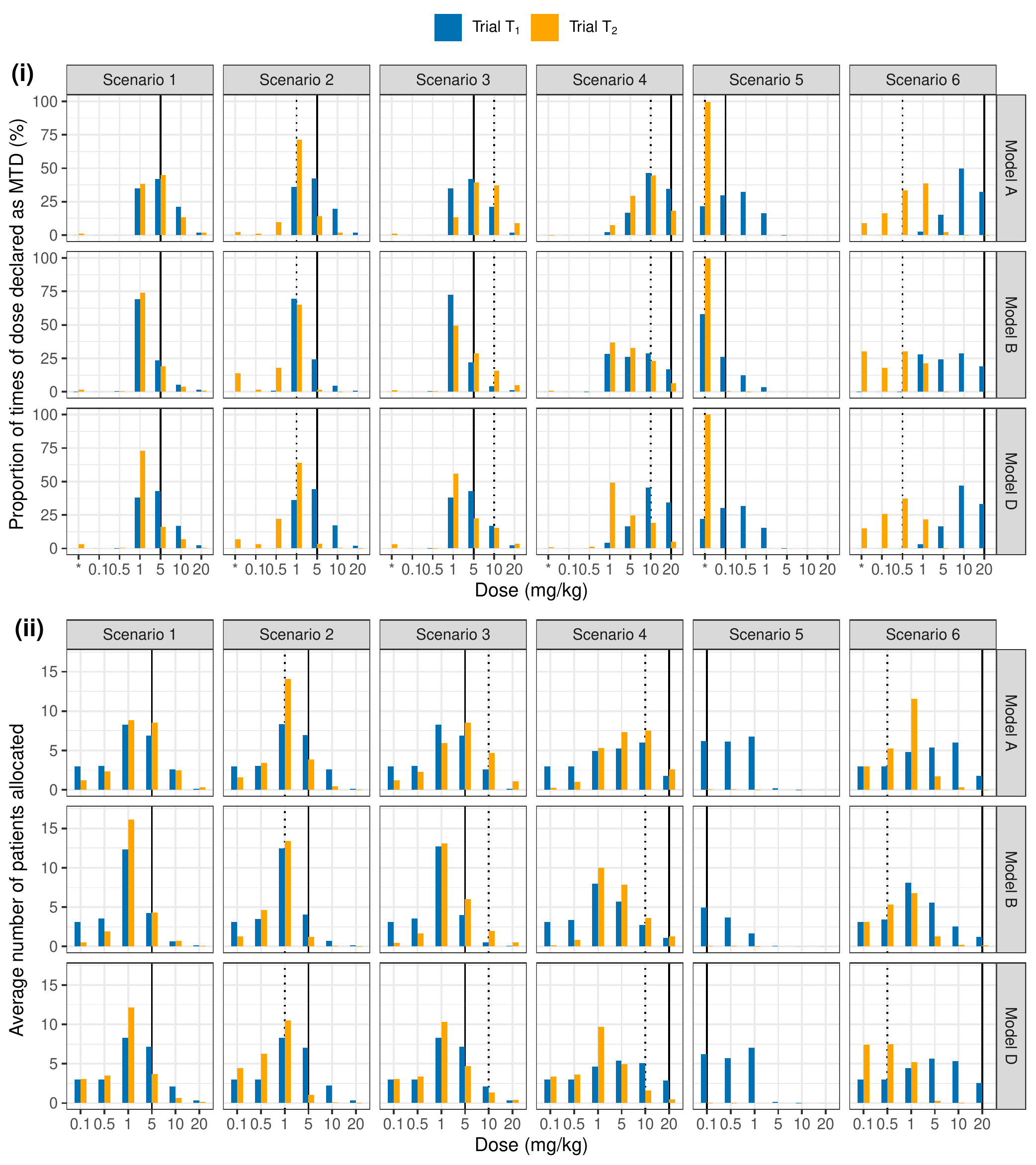}
\caption{Operating characteristics of the adaptive phase I dose-escalation trials in regions $\mathcal{R}_1$ and $\mathcal{R}_2$, conducted and analysed using Models A, B and D. The vertical black solid (dotted) line indicates the true MTD in the first-in-man trial $\mathcal{T}_1$ (trial $\mathcal{T}_2$) in each simulation scenario.}
\label{fig:PhIMetrics}
\end{figure}

[Figure \ref{fig:PhIMetrics} about here.] \\

Our primary interest lies in understanding the strength and weaknesses of the proposed Bayesian model, which leverages animal data as well as human toxicity data from a different subgroup. 
We focus here on the operating characteristics of Models A, B and D. 
Comparisons between Models A, C and E are presented in Figure S3 and the complete numerical results in Table S2 of the Supplementary Materials.
From Figure \ref{fig:PhIMetrics}, we see that Model A is able to beat the alternative analysis models across nearly all the simulation scenarios.
In scenarios 1 and 2, where animal data are highly predictive of the human DLT risk, compared with Model B, Models A and D lead to a higher percentage of correct selection (PCS) and a higher proportion of patients allocated to tolerable doses, with a DLT risks in the range [0.16, 0.33).
In scenario 1, DLT risks are identical across human subgroups.
Comparing Models A and D, allowing for information sharing across patient subgroups leads to an increase in the PCS for trial $\mathcal{T}_2$ from 16\% to 45\% in scenario 1.
Borrowing across human subgroups offers smaller but still meaningful gains in PCS in scenarios 2 and 3, when DLT risks are similar, but not identical, in the two regions.
Due to the ``no-skipping-of-dose'' restriction and a small sample size, it is challenging in scenario 4 to declare doses 20 and 10 mg/kg as the MTD in trials $\mathcal{T}_1$ and $\mathcal{T}_2$, respectively.
Nevertheless, trials performed using Model A have the highest PCS in regions $\mathcal{R}_1$ and $\mathcal{R}_2$. 
In particular, we see an increase of 25.3\% in PCS and, averagely, about 6 more patients treated at the true MTD of trial $\mathcal{T}_2$, comparing Model A with Model D in scenario 4.  \\

In scenario 5, all the Bayesian analysis models adequately protect patients from receiving overly toxic doses.
Due to the use of animal data, Models A and D tend to treat more patients than Model B with doses 0.5 and 1 mg/kg, which have human DLT risks exceeding 33\%.
However, the average number of patients experiencing a DLT is not substantially higher than the number under Model B.
Scenario 6 represents the case where the bridging assumption is incorrect. Comparing trial operating characteristics under Model A with those under Models C and D (which make no assumption of bridging), we find our approach can effectively discount information from trial $\mathcal{T}_1$ and so leads to a PCS which is comparable to that of Models C and D.
Nevertherless, Model A allocates more patients (specifically, 6 more on average) to dose 1 mg/kg than Models B and C in scenario 6.  \\

Comparing Model A with Model B in scenarios 1 -- 4, in which the bridging assumption is correct, there is a clear benefit of leveraging preclinical animal data and allowing for the possibility of non-exchangeability. In scenarios 1 -- 3, Model B assigned more patients of trial $\mathcal{T}_1$ to dose 1 mg/kg than the true MTD 5 mg/kg, due to the stated ``no-skipping'' dose-escalation constraint and our rule for defining the MTD: only the administered doses are eligible to be selected as a MTD.
Consequently, more trials were ended with a safer dose being declared as the MTD. 
In contrast, Models A and D which make use of animal data (of which the monkey data favour dose 5 mg/kg) helped faster escalation to the true MTD in these scenarios.
In scenarios 3 and 4, Model B experienced increasing difficulty distinguishing the region-specific MTD particularly for trial $\mathcal{T}_2$ when the true MTDs are on the top of set $\mathcal{D}_\ell$ and when differences between human toxicity across regions were small. 
Indeed, the assumption of full exchangeability led to excessive sharing of information between the two phase I clinical trials. 
In scenario 5, trial $\mathcal{T}_1$ conducted using Model B were more likely to be {\sl stopped early} for safety. 
Models A and B gave divergent operating characteristics in scenario 6. 
As Model B tends to underestimate the toxicity in region $\mathcal{R}_1$ in such a scenario (like scenario 4), too strong borrowing of information from $\mathcal{T}_1$ to $\mathcal{T}_2$ led to more $\mathcal{T}_2$ trials stopped early than Models A and D.
Focusing on Models A and D in scenario 6, using animal data provides advantages for correctly selecting the MTD in each region. Whereas, Model A allocated about 6 -- 7 more patients to dose 1 mg/kg in trial $\mathcal{T}_2$ and more often incorrectly select this dose as MTD, compared with Model D, as the loss of making an erroneous bridging assumption. \\

Referring to the Supplementary Materials, we can compare Model A with Models C and E in similar ways. 
In particular, the improved operating characteristics when comparing Models A and C should be interpreted as a mixture of benefit from using both animal data and appropriate bridging strategy.
The main disadvantage of Models C and E is that they are too extreme models for either not borrowing at all or completely pooling the trial data of $\mathcal{T}_1$ for trial $\mathcal{T}_2$.
We have also compared Models A -- E with respect to the posterior estimates of the dose-toxicity relationship in each human subgroup. Figure S4 of the Supplementary Materials show that Model A outperforms the others, providing very satisfactory characterisation of the association on the termination of a phase I clinical trial.  
We additionally ran simulations for a robust version of Model B with $w_{\ell R} = 0.20$.
Conclusions are similar with those written in \citet{SIBR:Neuenschwander2016} and \citet{SMMR:Zheng2019}, and thus will not be repeated in this paper.
 \\

\begin{figure}[!htb]
\centering
\captionsetup{font=scriptsize}
\includegraphics[width=0.75\linewidth]{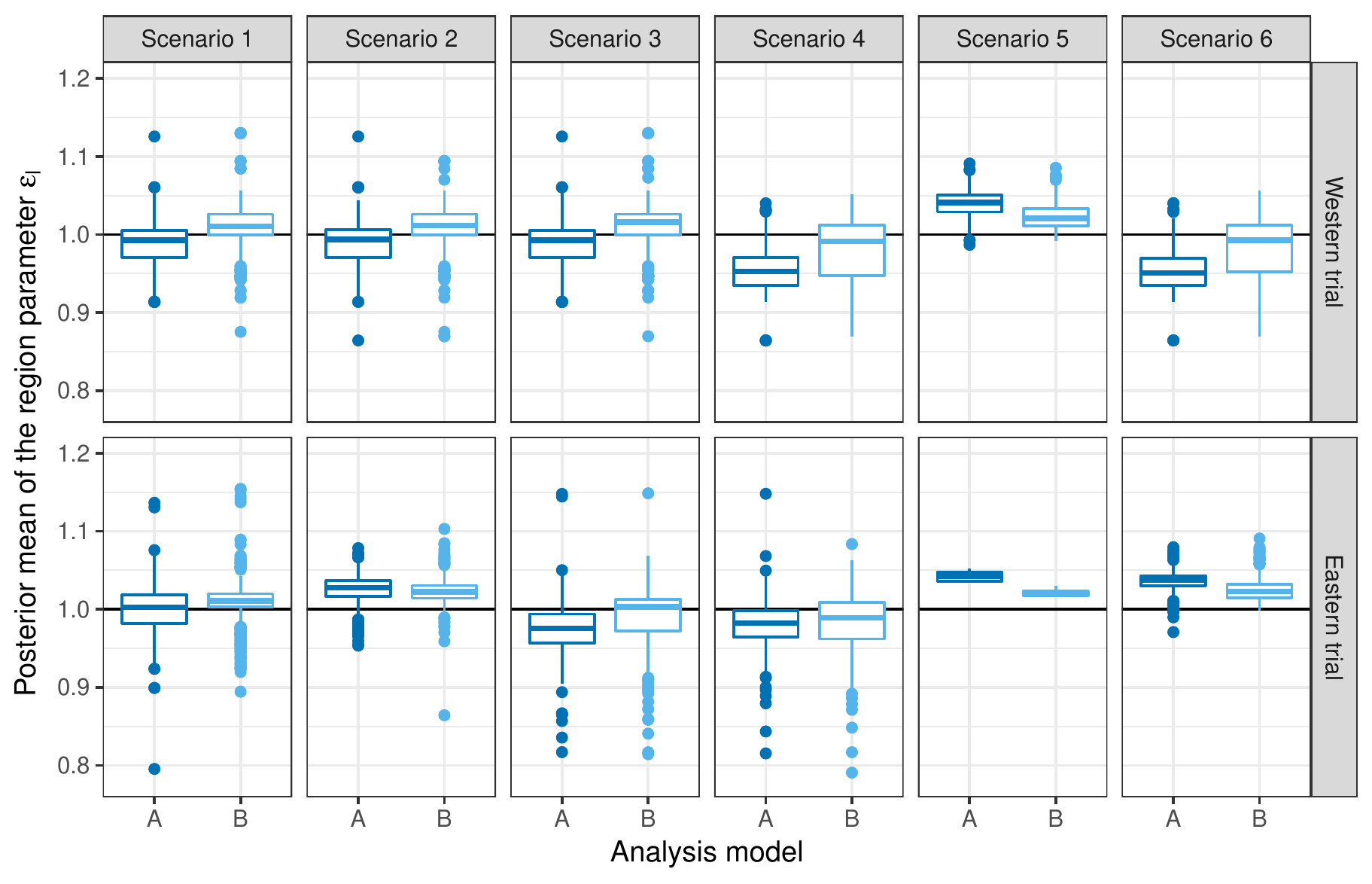}
\caption{Boxplots that depict the posterior means of the region parameter $\epsilon_\ell$ estimated by the end of {\sl completed} trials, designed using Model A or Model B. The horizontal black line represents the prior mean of $\epsilon_\ell$.}
\label{fig:RegionPar}
\end{figure}

[Figure \ref{fig:RegionPar} about here.] \\

We introduced a bridging parameter $\epsilon_\ell$ into the human dose-toxicity model in the form of \eqref{eq:Bdose-tox}.
Our approach (Model A) maps animal data onto the equivalent human dosing scale for the incorporation into human trials $\mathcal{T}_1$ and $\mathcal{T}_2$.
With inclusion of a random $\epsilon_\ell$ after setting $\delta_{\mathcal{A}_i} = 1$ for the human trials, the bridging parameter may also be suggestive of whether animal data over- or under-predict the toxicity of patients in a specific human subgroup. 
See Figure \ref{fig:RegionPar} for the boxplots of Model A in scenario 4, where animal data from both rat and monkey studies have underestimated the toxicity in human subjects.
We observe both $\epsilon_\ell$ for trials $\mathcal{T}_\ell, \, \ell = 1, 2$ shifted from the normal prior mean to take a value smaller than its normal prior mean which is 1, such that the random-effects distributions \eqref{eq:parmod} can explain some variability. The smaller difference between the values of $\epsilon_\ell$, the more variability will be explained by the human specific random-effects distribution BVN$(\boldsymbol{\mu}_\mathcal{H}, \Phi)$.
This can be perceived by the boxplots in scenarios 1 and 3 of Figure \ref{fig:RegionPar}. 
When the boxplot is shifted to take a value larger than 1 (the  normal prior mean), it suggests animal data are likely to have overestimated the human toxicity, see for example scenario 2 for trial $\mathcal{T}_2$.
As in scenario 5 many simulated trials will be stopped early for safety while $\epsilon_\ell$ will be estimated for a completed trial only, we will exclude the boxplots in this scenario for discussion. 
In contrast, the parameter $\epsilon_\ell$ embedded in Model B exclusively addresses the intrinsic differences arising from ethnicity between patient subgroups.
Within the same scenario, say, scenarios 1 and 3 where the bridging strategy should better be implemented, $\epsilon_\ell$ for both trials $\mathcal{T}_1$ and $\mathcal{T}_2$ take values centring around 1 (the normal prior mean). When the drug is more toxic to patients of regions $\mathcal{R}_1$ than $\mathcal{R}_2$, the boxplot of $\epsilon_\ell$ for trial $\mathcal{T}_2$ shifts upwards to take larger values, see the boxplots in scenarios 2 and 6.

\section{Discussion}
\label{sec:discuss}

Bridging studies have received considerable interest \citep{SMMR:Wadsworth2016}, as fewer resources may be needed to demonstrate drug behaviours by using relevant data from other subgroups, compared with the approach of establishing an independent package of clinical drug development. 
To date, methodology to extrapolate from foreign clinical data has been proposed mainly for phase II and phase III trials \citep{JBS:Hsiao2005, JBS:Chow2012, JBS:Tsou2012, JBS:Zhang2017}. 
Discussion was limited on making good use of co-data in phase I dose-escalation trials for informed decision making. Moreover, there appears to be scant literature on regarding preclinical animal data as a special kind of co-data for human trials. \\

In this paper, we have presented a robust Bayesian hierarchical model for leveraging co-data, available from both animal studies and phase I trials conducted in different patient subgroups, to improve the mid-course dose recommendations as well as the estimation of the subgroup-specific MTD. Our model can accommodate heterogeneity from various sources and lead to robust inferences. 
It down-weights co-data effectively in scenarios of inconsistency with the new trial data, and improves operating characteristics in scenarios of data consistency. 
For illustration, we presented several hypothetical data examples and the simulation study with two phase I trials, assuming that they are planned in different geographic regions and conducted one after the other. 
The proposed Bayesian approach, however, works equally well if the phase I trials would have been run concurrently.
In this case, each trial will be regarded as a stratum and the interim dose recommendations will be based on accmulating data from the ongoing trials. 
We also note there is no restriction on the number of phase I trials, for which we wish to base the decision making on the co-data.
Potentially, a large number of human subgroups can be advantageous for estimating the between-trial heterogeneity, and therefore better determine the degree of borrowing in each human trial $\mathcal{T}_\ell, \, \ell = 1, \dots, L$.\\

Our Bayesian model presented in this paper is highly relevant with the development of early phase basket trials \citep{AO:Renfro2017, BIOS:Psioda2019}, where robust hierarchical models have been considered for borrowing of information \citep{PS:Neuenschwander2016, CT:Chu2018}. 
Since our approach can bridge across patient subgroups, it has the potential to be applied for analysing basket trials.
Historical data would have to be carefully chosen to formulate the exchangeability distributions.
It would also be worth refining the borrowing based on how similar the outcome data would be across subgroups, drawing ideas from\citep{BIOS:ZhengWason2019}.

\newpage

\appendix

\section{OpenBUGS code for implemention}
\label{sec:appdCode}

\begin{verbatim}
model{
	# sampling model
	# Mdoses: total number of doses tested in animal species
	for(j in 1:MdoseA){
	linA[j] <- theta[StudyA[j], 1] 
			+ exp(theta[StudyA[j], 2])*log(deltaA[Species[j]]*DoseA[j]/DoseRef)
	logit(pToxA[j]) <- linA[j]
	NtoxA[j] ~ dbin(pToxA[j], NsubA[j])
	}

	zero[1] <- 0
	zero[2] <- 0

	# theta = (theta[i, 1], theta[i, 2]) derived from each animal study are ready for the use
	# on the equivalent human dosing scale
	for(i in 1:NstudyA){
		for(j in 1:MdoseH){
			lin[i, j] <- theta[i, 1] + exp(theta[i, 2])*log(DoseH[j]/DoseRef)
		}

		# sp.ind[i]: index function to specify
		# which animal species the Study i belongs to
		theta[i, 1] <- mu.sp[sp.ind[i], 1] + re.A[i, 1]
		theta[i, 2] <- mu.sp[sp.ind[i], 2] + re.A[i, 2]
		re.A[i, 1:2] ~ dmnorm(zero[1:2], prec.Psi[1:2, 1:2])

			# PInd[]: matrice of the trivial/non-trivial weights
			# trivial weights for animals means 0 prob of NEX
			# to assure theta_i are fully exchangeable withing the same species
			sp.ind[i] ~ dcat(PInd[i, 1:n.sp])
	}

	# the K EX distributions based upon animal species clusters
	for(k in 1:n.sp){
		deltaA[k] <- exp(Prior.mn.delta[k] + Prior.sd.delta[k]*log.delta01[k])
		log.delta01[k] ~ dnorm(0, 1)
		mu.sp[k, 1] <- muA[1] + re.m[k, 1]
		mu.sp[k, 2] <- muA[2] + re.m[k, 2]
		re.m[k, 1:2] ~ dmnorm(zero[1:2], prec.Sigma[1:2, 1:2])

		theta.predH[k, 1] <- mu.sp[k, 1] + re.h[k, 1]
		theta.predH[k, 2] <- mu.sp[k, 2] + re.h[k, 2]
		re.h[k, 1:2] ~ dmnorm(zero[1:2], prec.Psi[1:2, 1:2])
	}


	for(i in 1:n.sb){

		for(k in 1:n.sp){
			mix.theta[i, k, 1] <- theta.predH[k, 1] 
			mix.theta[i, k, 2] <- theta.predH[k, 2]
		}

			mix.theta[i, (n.sp+1), 1] <- muH[1] + re.s[i, 1]
			mix.theta[i, (n.sp+1), 2] <- muH[2] + re.s[i, 2]
			re.s[i, 1:2] ~ dmnorm(zero[1:2], prec.Phi[1:2, 1:2])

			mix.theta[i, (n.sp+2), 1:2] ~ dmnorm(Prior.mw[1:2], prec.sw[1:2, 1:2])

		# pick theta
		theta.star[i, 1] <- mix.theta[i, exch.ind[i], 1]
		theta.star[i, 2] <- mix.theta[i, exch.ind[i], 2]

		# latent mixture indicators:
		exch.ind[i] ~ dcat(wMix[i, 1:(n.sp+2)])	
			for(ii in 1:(n.sp+2)){
				each[i, ii] <- equals(exch.ind[i], ii)
			}

		# Update theta.star[i, 1:2] using the phase I trial data from various subgroups
		for(j in 1:MdoseH){
			linH[i, j] <- theta.star[i, 1] 
					+ exp(theta.star[i, 2])*log(epsilonH[i]*DoseH[j]/DoseRef)
			logit(pToxH[i, j]) <- linH[i, j]
			NtoxH[i, j] ~ dbin(pToxH[i, j], NsubH[i, j])

			pCat[i, j, 1] <- step(pTox.cut[1] - pToxH[i, j])
			pCat[i, j, 2] <- step(pTox.cut[2] - pToxH[i, j]) 
						- step(pTox.cut[1] - pToxH[i, j])
			pCat[i, j, 3] <- step(1 - pToxH[i, j]) 
						- step(pTox.cut[2] - pToxH[i, j])
		}
			epsilonH[i] <- Prior.mn.epsilon[i] + Prior.sd.epsilon[i]*epsilon01[i]
			epsilon01[i] ~ dnorm(0, 1)I(-3.921, 3.921)
	}

# Hyperpriors for the human-specific population means muH[1:2]
muH[1] ~ dnorm(Prior.mH1[1], prec.mH1)I(-10, 10)
muH[2] ~ dnorm(Prior.mH2[1], prec.mH2)I(-5, 5)

prec.mH1 <- pow(Prior.mH1[2], -2)
prec.mH2 <- pow(Prior.mH2[2], -2)

# Hyperpriors for the `supra-spiece' population means muA[1:2] 
muA[1] ~ dnorm(Prior.mA1[1], prec.mA1)I(-10, 10)
muA[2] ~ dnorm(Prior.mA2[1], prec.mA2)I(-5, 5)

prec.mA1 <- pow(Prior.mA1[2], -2)
prec.mA2 <- pow(Prior.mA2[2], -2)

# Hyperpriors for the covariance matrix, prec.Psi[1:2, 1:2]
prec.tau1 <- pow(Prior.tau.HN[1], -2)
prec.tau2 <- pow(Prior.tau.HN[2], -2)
tauA[1] ~ dnorm(0, prec.tau1)I(0.001,)
tauA[2] ~ dnorm(0, prec.tau2)I(0.001,)

covA.ex[1, 1] <- pow(tauA[1], 2)
covA.ex[2, 2] <- pow(tauA[2], 2)
covA.ex[1, 2] <- tauA[1]*tauA[2]*rhoA
covA.ex[2, 1] <- covA.ex[1, 2]
prec.Psi[1:2, 1:2] <- inverse(covA.ex[1:2, 1:2])

	rhoA ~ dunif(Prior.rho[1], Prior.rho[2])

# Hyperpriors for the covariance matrix, prec.Sigma[1:2, 1:2]
prec.sigma1 <- pow(Prior.sigma.HN[1], -2)
prec.sigma2 <- pow(Prior.sigma.HN[2], -2)
sigma[1] ~ dnorm(0, prec.sigma1)I(0.001,)
sigma[2] ~ dnorm(0, prec.sigma2)I(0.001,)
covA.sig[1, 1] <- pow(sigma[1], 2)
covA.sig[2, 2] <- pow(sigma[2], 2)
covA.sig[1, 2] <- sigma[1]*sigma[2]*kappaA
covA.sig[2, 1] <- covA.sig[1, 2]
prec.Sigma[1:2, 1:2] <- inverse(covA.sig[1:2, 1:2])

	kappaA ~ dunif(Prior.kappa[1], Prior.kappa[2])

# Hyperpriors for the covariance matrix, prec.Phi[1:2, 1:2]
prec.tau3 <- pow(Prior.tau.HN[3], -2)
prec.tau4 <- pow(Prior.tau.HN[4], -2)
tauH[1] ~ dnorm(0, prec.tau3)I(0.001,)
tauH[2] ~ dnorm(0, prec.tau4)I(0.001,)

covH.ex[1, 1] <- pow(tauH[1], 2)
covH.ex[2, 2] <- pow(tauH[2], 2)
covH.ex[1, 2] <- tauH[1]*tauH[2]*rhoH
covH.ex[2, 1] <- covH.ex[1, 2]
prec.Phi[1:2, 1:2] <- inverse(covH.ex[1:2, 1:2])

	rhoH ~ dunif(Prior.rho[1], Prior.rho[2])

# Weakly-informative hyperpriors for the covariance matrix, prec.sw[1:2, 1:2]
cov.rb[1, 1] <- pow(Prior.sw[1], 2)
cov.rb[2, 2] <- pow(Prior.sw[2], 2)
cov.rb[1, 2] <- Prior.sw[1]*Prior.sw[2]*Prior.corr
cov.rb[2, 1] <- cov.rb[1, 2]
prec.sw[1:2, 1:2] <- inverse(cov.rb[1:2, 1:2])
}
\end{verbatim}

\newpage

\bibliographystyle{apalike}

\begin{thebibliography}{}


\bibitem[Chow et~al., 2012]{JBS:Chow2012}
Chow, S.-C., Chiang, C., pei Liu, J., and Hsiao, C.-F. (2012).
\newblock Statistical methods for bridging studies.
\newblock {\em Journal of Biopharmaceutical Statistics}, 22(5):903--915.

\bibitem[Chu and Yuan, 2018]{CT:Chu2018}
Chu, Y. and Yuan, Y. (2018).
\newblock A {Bayesian} basket trial design using a calibrated {Bayesian}
  hierarchical model.
\newblock {\em Clinical Trials}, 15(2):149--158.

\bibitem[de~Haen, 1975]{DICP:Haen1975}
de~Haen, P. (1975).
\newblock The drug lag-does it exist in europe?
\newblock {\em Drug Intelligence \& Clinical Pharmacy}, 9(3):144--150.

\bibitem[Gelman et~al., 2008]{AAS:Gelman2008}
Gelman, A., Jakulin, A., Pittau, M., and Su, Y. (2008).
\newblock A weakly informative default prior distribution for logistic and
  other regression models.
\newblock {\em The Annals of Applied Statistics}, 2(4):1360--1383.

\bibitem[GlaxoSmithKline, 2016]{GSK:NCT02671188}
GlaxoSmithKline (2016).
\newblock A study to evaluate the safety, mode of action and clinical efficacy
  of {GSK3050002} in subjects with psoriatic arthritis.
\newblock {\em Bethesda (MD): National Library of Medicine (US)}.
\newblock (Available from: https://clinicaltrials.gov/ct2/show/NCT02671188 (NLM
  Identifier: NCT02671188)) [Last accessed in September 2018].

\bibitem[GlaxoSmithKline, 2018]{GSK:PhIprotocol}
GlaxoSmithKline (2018).
\newblock A phase 1, randomized, double-blind (sponsor open),
  placebo-controlled, single dose escalation trial to evaluate the safety,
  tolerability pharmacokinetics and pharmacodynamics of {GSK3050002
  (anti-CCL20} monoclonal antibody) in healthy male volunteers.
\newblock {\em GlaxoSmithKline Research \& Development Limited}.
\newblock (Available from:
  https://www.gsk-clinicalstudyregister.com/files2/gsk-200784-protocol-redact.pdf
  [Last accessed in September 2018].

\bibitem[Hsiao et~al., 2004]{JBS:Hsiao2005}
Hsiao, C.-F., Xu, J.-Z., and pei Liu, J. (2004).
\newblock A two-stage design for bridging studies.
\newblock {\em Journal of Biopharmaceutical Statistics}, 15(1):75--83.

\bibitem[Huang et~al., 2012]{JBS:Huang2012}
Huang, Q., Chen, G., Yuan, Z., and Lan, K. K.~G. (2012).
\newblock Design and sample size considerations for simultaneous global drug
  development program.
\newblock {\em Journal of Biopharmaceutical Statistics}, 22(5):1060--1073.

\bibitem[{(ICH) E5 Guideline}, 1998]{EMA:ICH1998}
{(ICH) E5 Guideline} (1998).
\newblock {\em Tripartite Guidance E5 (R1), Ethnic Factors in the Acceptability
  of Foreign Clinical Data}.
\newblock European Medicines Agency: London, E14 4HB, UK.

\bibitem[{(ICH) E5 Guideline}, 2006]{EMA:ICH2006}
{(ICH) E5 Guideline} (2006).
\newblock {\em Question and Answers for the ICH E5 Guideline on Ethnic Factors
  in the Acceptability of Foreign Data}.
\newblock European Medicines Agency: London, E14 4HB, UK.

\bibitem[Kamrin, 1988]{Kamrin1988}
Kamrin, M. (1988).
\newblock {\em Toxicology - A Primer on Toxicology Principles and
  Applications}.
\newblock CRC Press.

\bibitem[Li and Wang, 2012]{JBS:Li2012}
Li, N. and Wang, W. (2012).
\newblock Practical and statistical considerations on simultaneous global drug
  development.
\newblock {\em Journal of Biopharmaceutical Statistics}, 22(5):1074--1077.

\bibitem[Liu et~al., 2015]{SiM:Liu2015}
Liu, S., Pan, H., Xia, J., Huang, Q., and Yuan, Y. (2015).
\newblock Bridging continual reassessment method for phase {I} clinical trials
  in different ethnic populations.
\newblock {\em Statistics in Medicine}, 34(10):1681--1694.

\bibitem[Lunn et~al., 2009]{SiM:Lunn2009}
Lunn, D., Spiegelhalter, D., Thomas, A., and Best, N. (2009).
\newblock The bugs project: Evolution, critique and future directions.
\newblock {\em Statistics in Medicine}, 28(25):3049--3067.

\bibitem[Mizugaki et~al., 2015]{JCO:Mizugaki2015}
Mizugaki, H., Yamamoto, N., Fujiwara, Y., Nokihara, H., Yamada, Y., and Tamura,
  T. (2015).
\newblock Current status of single-agent phase i trials in japan: Toward
  globalization.
\newblock {\em Journal of Clinical Oncology}, 33(18):2051--2061.

\bibitem[Morita et~al., 2008]{BIOM:Morita2008}
Morita, S., Thall, P.~F., and M\"{u}ller, P. (2008).
\newblock Determining the effective sample size of a parametric prior.
\newblock {\em Biometrics}, 64(2):595--602.

\bibitem[Neuenschwander et~al., 2008]{SIM:Neuenschwander2008}
Neuenschwander, B., Branson, M., and Gsponer, T. (2008).
\newblock Critical aspects of the {Bayesian} approach to phase {I} cancer
  trials.
\newblock {\em Statistics in Medicine}, 27(13):2420--2439.

\bibitem[Neuenschwander et~al., 2016a]{SIBR:Neuenschwander2016}
Neuenschwander, B., Roychoudhury, S., and Schmidli, H. (2016a).
\newblock On the use of co-data in clinical trials.
\newblock {\em Statistics in Biopharmaceutical Research}, 8(3):345--354.

\bibitem[Neuenschwander et~al., 2016b]{PS:Neuenschwander2016}
Neuenschwander, B., Wandel, S., Roychoudhury, S., and Bailey, S. (2016b).
\newblock Robust exchangeability designs for early phase clinical trials with
  multiple strata.
\newblock {\em Pharmaceutical Statistics}, 15(2):123--134.

\bibitem[Ogura et~al., 2014]{TIRS:Ogura2014}
Ogura, T., Morita, S., Yonemori, K., Nonaka, T., and Urano, T. (2014).
\newblock Exploring ethnic differences in toxicity in early-phase clinical
  trials for oncology drugs.
\newblock {\em Therapeutic Innovation \& Regulatory Science}, 48(5):644--650.

\bibitem[O'Quigley and Iasonos, 2014]{SBR:Quigley2014}
O'Quigley, J. and Iasonos, A. (2014).
\newblock Bridging solutions in dose finding problems.
\newblock {\em Statistics in biopharmaceutical research}, 6(2):185—197.

\bibitem[O'Quigley et~al., 1999]{JBS:Quigley1999}
O'Quigley, J., Shen, L.~Z., and Gamst, A. (1999).
\newblock Two-sample continual reassessment method.
\newblock {\em Journal of Biopharmaceutical Statistics}, 9(1):17--44.

\bibitem[PMDA, 2007]{JPMA:Principles2007}
PMDA (2007).
\newblock {\em Basic Principles on Global Clinical Trials, Notification No.
  0928010}.
\newblock Ministry of Health, Labour and Welfare: Tokyo, Japan.
\newblock This document is an informal translation by PMDA of the final
  notification published in Japanese on Sep. 28th 2007 and is intended to use
  as a reference for considering global clinical trials. Last access:
  September, 2018.

\bibitem[Psioda et~al., 2019]{BIOS:Psioda2019}
Psioda, M., Xu, J., Jiang, Q., Ke, C., Yang, Z., and Ibrahim, J. (2019).
\newblock {Bayesian adaptive basket trial design using model averaging}.
\newblock {\em Biostatistics}.
\newblock Epub ahead of print.

\bibitem[{R Core Team}, 2017]{RSoftware}
{R Core Team} (2017).
\newblock R: A language and environment for statistical computing.
\newblock {\em R Foundation for Statistical Computing}.
\newblock Available online at https://www.R-project.org/.

\bibitem[Renfro and Sargent, 2017]{AO:Renfro2017}
Renfro, L. and Sargent, D. (2017).
\newblock Statistical controversies in clinical research: basket trials,
  umbrella trials, and other master protocols: a review and examples.
\newblock {\em Annals of Oncology}, 28(1):34--43.

\bibitem[Takeda and Morita, 2018]{PS:Takeda2018}
Takeda, K. and Morita, S. (2018).
\newblock Bayesian dose-finding phase {I} trial design incorporating historical
  data from a preceding trial.
\newblock {\em Pharmaceutical Statistics}, 17(4):372--382.

\bibitem[Thomas, 2017]{R2OpenBUGS}
Thomas, N. (2017).
\newblock R2openbugs: Running openbugs from r.
\newblock {\em CRAN}.
\newblock R package version 3.2-3.2.

\bibitem[Tsong, 2012]{JBS:Tsong2012}
Tsong, Y. (2012).
\newblock Statistical considerations on design and analysis of bridging and
  multiregional clinical trials.
\newblock {\em Journal of Biopharmaceutical Statistics}, 22(5):1078--1080.

\bibitem[Tsou et~al., 2012]{JBS:Tsou2012}
Tsou, H.-H., Tsong, Y., Liu, J.-T., Dong, X., and Wu, Y. (2012).
\newblock Weighted evidence approach of bridging study.
\newblock {\em Journal of Biopharmaceutical Statistics}, 22(5):952--965.

\bibitem[Ueno et~al., 2013]{CPT:Ueno2013}
Ueno, T., Asahina, Y., Tanaka, A., Yamada, H., Nakamura, M., and Uyama, Y.
  (2013).
\newblock Significant differences in drug lag in clinical development among
  various strategies used for regulatory submissions in japan.
\newblock {\em Clinical Pharmacology \& Therapeutics}, 95(5):533--541.

\bibitem[USFDA, 2005]{FDA:FIH2005}
USFDA (2005).
\newblock {\em Estimating the maximum safe starting dose in initial clinical
  trials for therapeutics in adult healthy volunteers}.
\newblock US Food and Drug Administration: Rockville, MD.

\bibitem[Viergever and Li, 2015]{BMJ:Viergever2015}
Viergever, R.~F. and Li, K. (2015).
\newblock Trends in global clinical trial registration: an analysis of numbers
  of registered clinical trials in different parts of the world from 2004 to
  2013.
\newblock {\em BMJ Open}, 5(9).

\bibitem[Wadsworth et~al., 2018]{SMMR:Wadsworth2016}
Wadsworth, I., Hampson, L., and Jaki, T. (2018).
\newblock Extrapolation of efficacy and other data to support the development
  of new medicines for children: a systematic review of methods.
\newblock {\em Statistical Methods in Medical Research}, 27(2):398--413.

\bibitem[Wages et~al., 2015]{PS:Wages2015}
Wages, N.~A., Read, P.~W., and Petroni, G.~R. (2015).
\newblock A phase i/ii adaptive design for heterogeneous groups with
  application to a stereotactic body radiation therapy trial.
\newblock {\em Pharmaceutical Statistics}, 14(4):302--310.

\bibitem[Whitehead and Williamson, 1998]{BioPS:Whitehead1998}
Whitehead, J. and Williamson, D. (1998).
\newblock Bayesian decision procedures based on logistic regression models for
  dose-finding studies.
\newblock {\em Journal of Biopharmaceutical Statistics}, 8(3):445--467.
\newblock PMID: 9741859.

\bibitem[Wileman and Mishra, 2010]{PCR:Wileman2010}
Wileman, H. and Mishra, A. (2010).
\newblock Drug lag and key regulatory barriers in the emerging markets.
\newblock {\em Perspectives in Clinical Research}, 1(2):51--56.

\bibitem[Yin and Yuan, 2009]{JASA:Yin2009}
Yin, G. and Yuan, Y. (2009).
\newblock Bayesian model averaging continual reassessment method in phase {I}
  clinical trials.
\newblock {\em Journal of the American Statistical Association},
  104(487):954--968.

\bibitem[Zhang et~al., 2017]{JBS:Zhang2017}
Zhang, T., Lipkovich, I., and Marchenko, O. (2017).
\newblock Bridging data across studies using frequentist and {Bayesian}
  estimation.
\newblock {\em Journal of Biopharmaceutical Statistics}, 27(3):426--441.

\bibitem[Zheng and Hampson, 2019]{BJ:Zheng2019}
Zheng, H. and Hampson, L.~V. (2019).
\newblock A {Bayesian} decision-theoretic approach to incorporate preclinical
  information into phase {I} oncology trials.
\newblock {\em arXiv preprint}, pages 1--34.
\newblock arXiv:1907.13620.

\bibitem[Zheng et~al., 2019]{SMMR:Zheng2019}
Zheng, H., Hampson, L.~V., and Wandel, S. (2019).
\newblock A robust {Bayesian} meta-analytic approach to incorporate animal data
  into phase {I} oncology trials.
\newblock {\em Statistical Methods in Medical Research}, pages 1--17.
\newblock Epub ahead of print.

\bibitem[Zheng and Wason, 2019]{BIOS:ZhengWason2019}
Zheng, H. and Wason, J.~M. (2019).
\newblock Borrowing of information across patient subgroups in a basket trial
  based on distributional discrepancy.
\newblock {\em arXiv preprint}, pages 1--19.
\newblock arXiv:1908.05091.


\end{thebibliography}


\clearpage

\begin{table}[!htb]
\scriptsize
\centering
\captionsetup{font=scriptsize}
\caption{Effective sample sizes of the marginal predictive posteriors (priors) for the DLT risk per dose, on the completion of trial $\mathcal{T}_1$ (start of trial $\mathcal{T}_2$), given the $\mathcal{T}_1$ trial data simulated from (i) a divergent scenario and (ii) a consistent scenario, respectively}
\vspace{-1em}
\begin{tabular}{@{}llccccccccccccc@{}}
\toprule
& &\multicolumn{6}{@{}c}{Trial $\mathcal{T}_1$}  & &\multicolumn{6}{@{}c}{Trial $\mathcal{T}_2$} \\
\cline{3-8} \cline{10-15}
& &$d_{\ell 1}$  &$d_{\ell 2}$  &$d_{\ell 3}$  &$d_{\ell 4}$  &$d_{\ell 5}$  &$d_{\ell 6}$  &  &$d_{\ell 1}$  &$d_{\ell 2}$  &$d_{\ell 3}$  &$d_{\ell 4}$  &$d_{\ell 5}$  &$d_{\ell 6}$   \\
& & 0.1            & 0.5           & 1            & 5           & 10            & 20      &      & 0.1            & 0.5           & 1            & 5           & 10            & 20     \\
\midrule
Sc (i) & Posterior/Prior means     & 0.028  & 0.046  & 0.059  & 0.107  & 0.144  & 0.199  &  & 0.070  & 0.114  & 0.144  & 0.273  & 0.361  & 0.438  \\
& Posterior/Prior std dev.  & 0.032  & 0.043  & 0.049  & 0.068  & 0.079  & 0.105  &  & 0.110  & 0.141  & 0.158  & 0.221  & 0.260  & 0.281   \\
\cmidrule{2-15}
& ESS          & 25.1   & 23.1   & 22.1   & 19.8   & 18.7   & 13.5  &     
             & 4.4   & 4.1   & 4.0   & 3.0   & 2.4   & 2.1  \\
& \quad $a$    & 0.7  & 1.1  & 1.3  & 2.1  & 2.7  & 2.7   &   & 0.3  & 0.5  & 0.6  & 0.8  & 0.9 & 0.9 \\
& \quad $b$    & 24.4  & 22.0  & 20.8  & 17.7  & 16.0  & 10.8   &  & 4.1   & 3.6   & 3.4   & 2.2   & 1.5   & 1.2   \\

\midrule
Sc (ii) & Posterior/Prior means     & 0.047   & 0.092   & 0.128   & 0.315   & 0.418   & 0.508  &  & 0.074   & 0.123   & 0.156   & 0.300   & 0.396   & 0.483  \\
& Posterior/Prior std dev.  & 0.043   & 0.058   & 0.065   & 0.154   & 0.207   & 0.229  &  & 0.110  & 0.138  & 0.153  & 0.206  & 0.240  & 0.256   \\

\cmidrule{2-15}
& ESS          & 23.0  & 23.4  & 25.8   & 8.0   & 4.7   & 3.7  &
              & 4.7   & 4.7   & 4.6   & 4.0   & 3.1   & 2.8  \\
& \quad $a$    & 1.1  & 2.1  & 3.3  & 2.5  & 2.0  & 1.9   &   
                & 0.3  & 0.6  & 0.7  & 1.2  & 1.2  & 1.4 \\
& \quad $b$    & 21.9  & 21.3  & 22.5   & 5.5   & 2.7   & 1.8   &  
                & 4.4   & 4.1   & 3.9   & 2.8   & 1.9   & 1.4   \\

\bottomrule
\end{tabular}
\label{tab:pseudoptsT2}
\vspace{-0.8em}
\end{table}

\begin{table}[!htb]
\scriptsize
\centering
\captionsetup{font=scriptsize}
\caption{Simulation scenarios for the true probability of toxicity in humans for the phase I trials $\mathcal{T}_1$ and $\mathcal{T}_2$. The figure in bold indicates the target dose closest to the true MTD in each region.}
\vspace{-1.2em}
\begin{tabular}{@{}lccccccccccccc@{}}
\toprule
&\multicolumn{6}{@{}c}{Trial $\mathcal{T}_1$}  & &\multicolumn{6}{@{}c}{Trial $\mathcal{T}_2$} \\
\cline{2-7} \cline{9-14}
&$d_{\ell 1}$  &$d_{\ell 2}$  &$d_{\ell 3}$  &$d_{\ell 4}$  &$d_{\ell 5}$  &$d_{\ell 6}$  &  &$d_{\ell 1}$  &$d_{\ell 2}$  &$d_{\ell 3}$  &$d_{\ell 4}$  &$d_{\ell 5}$  &$d_{\ell 6}$    \\
& 0.1            & 0.5           & 1            & 5           & 10            & 20      &      & 0.1            & 0.5           & 1            & 5           & 10            & 20     \\
\midrule
Scenario 1     & 0.01    & 0.03    & 0.10    & \bf{0.25}    & 0.34    & 0.47    &  
               & 0.01    & 0.03    & 0.10    & \bf{0.25}    & 0.34    & 0.47  \\
Scenario 2     & 0.01    & 0.03    & 0.10    & \bf{0.25}    & 0.34    & 0.47    &
               & 0.05    & 0.12    & \bf{0.25}    & 0.37    & 0.50    & 0.60   \\
Scenario 3     & 0.01    & 0.03    & 0.10    & \bf{0.25}    & 0.34    & 0.47    &
               & 0.01    & 0.03    & 0.07    & 0.15    & \bf{0.25}    & 0.37   \\
Scenario 4     & 0.01    & 0.03    & 0.05    & 0.08    & 0.15    & \bf{0.25}    &
               & 0.02    & 0.05    & 0.07    & 0.12    & \bf{0.25}    & 0.36   \\  
Scenario 5     & \bf{0.25}    & 0.34    & 0.47    & 0.55    & 0.65    & 0.75    &
               & 0.40    & 0.50    & 0.60    & 0.70    & 0.80    & 0.90   \\       
Scenario 6     & 0.01    & 0.03    & 0.05    & 0.08    & 0.15    & \bf{0.25}    &
               & 0.10    & \bf{0.25}    & 0.36    & 0.50    & 0.60    & 0.68   \\                 
\bottomrule
\end{tabular}
\label{tab:pToxH}
\vspace{-1em}
\end{table}

\clearpage


\clearpage



\section*{Supplementary material}

# Web-based Supplementary Materials for:
# A Bayesian hierarchical model for bridging across patient subgroups in phase I clinical trials with animal data

**by Haiyan Zheng, Lisa V. Hampson, Thomas Jaki**

### A. THE SIMULATED ANIMAL DATA AND PREDICTIVE PRIORS FOR HUMAN TOXICITY

Figure S1 represent the simulated animal data used for the numerical studies. The height of the bar represents the number of animal subjects treated, and the height of the dark grey segment counts the number of toxicity. Doses listed in brown are those administered to either rats and monkeys, which are translated onto an equivalent human dosing scale in black. Projections are made by scaling animal doses using the prior median of $\delta_{\text{Rat}}$ or $\delta_{\text{Monkey}}$.

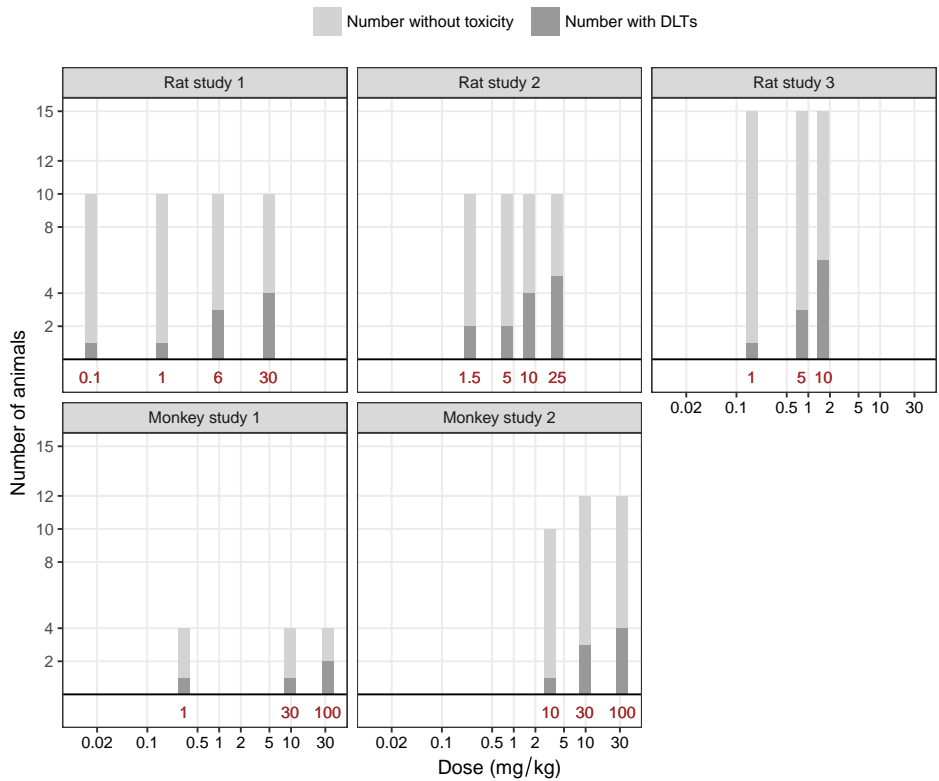

Figure S1: Hypothetical preclinical data in rats and monkeys, simulated for the illustrative examples as well as the simulation study.

With the simulated animal data and the priors specified in Section 3.1, we further obtain the marginal predictive priors that are shown graphically in Figure S2. We approximate these marginal predictive priors with a series of Beta$(a, b)$ distributions and report the prior effective sample size (ESS) for the predicted human toxicity at each dose in Table S1.

1

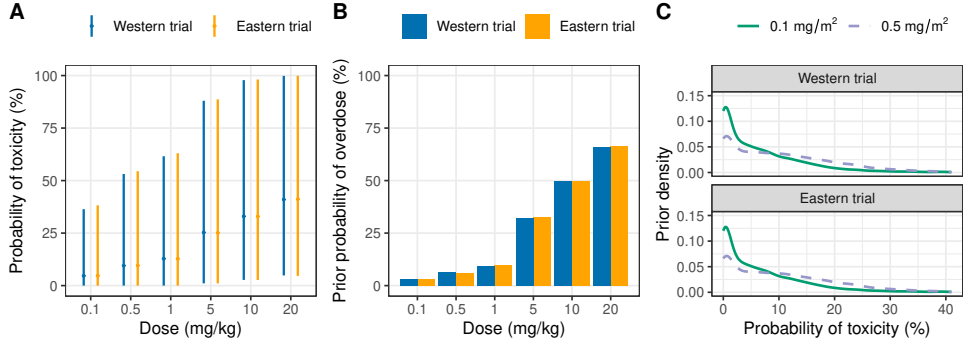

Figure S2: Summaries about the robust, marginal predictive priors for human toxicity based on the animal data. Panel A shows median and 95% credible interval of the marginal predictive prior for human toxicity at each dose. Panel B presents the prior interval probability of overdose, and Panel C displays prior densities for the risks of toxicity at potential starting doses.

Table S1: Effective sample sizes of the marginal predictive priors for the DLT risk per dose, obtained before trial $\mathcal{T}_1$ begins by fitting the robust Bayesian model, which translates the animal data onto an equivalent human dosing scale.

|  | Trial $\mathcal{T}_1$ | | | | | | Trial $\mathcal{T}_2$ | | | | | |
| --- | --- | --- | --- | --- | --- | --- | --- | --- | --- | --- | --- | --- |
|  | $d_{\ell 1}$ 0.1 | $d_{\ell 2}$ 0.5 | $d_{\ell 3}$ 1 | $d_{\ell 4}$ 5 | $d_{\ell 5}$ 10 | $d_{\ell 6}$ 20 | $d_{\ell 1}$ 0.1 | $d_{\ell 2}$ 0.5 | $d_{\ell 3}$ 1 | $d_{\ell 4}$ 5 | $d_{\ell 5}$ 10 | $d_{\ell 6}$ 20 |
| Prior means | 0.093 | 0.148 | 0.182 | 0.302 | 0.374 | 0.444 | 0.093 | 0.148 | 0.182 | 0.302 | 0.374 | 0.444 |
| Prior std dev. | 0.096 | 0.120 | 0.133 | 0.172 | 0.195 | 0.211 | 0.096 | 0.120 | 0.133 | 0.172 | 0.195 | 0.211 |
| ESS | 8.2 | 7.7 | 7.4 | 6.2 | 5.1 | 4.5 | 8.2 | 7.7 | 7.4 | 6.2 | 5.1 | 4.5 |
| $a$ | 0.8 | 1.1 | 1.3 | 1.9 | 1.9 | 2.0 | 0.8 | 1.1 | 1.3 | 1.9 | 1.9 | 2.0 |
| $b$ | 7.4 | 6.6 | 6.1 | 4.3 | 3.2 | 2.5 | 7.4 | 6.6 | 6.1 | 4.3 | 3.2 | 2.5 |

## B. SIMULATION RESULTS THAT ARE NOT PRESENTED IN THE MAIN PAPER

In the main mauscript, we have compared Model A (the proposed approach) with Models B and D and interpreted the results in terms of benefit attributed to either using the robust bridging strategy (Model A versus Model B) or animal data (comparing Model A and D). Here, we present additional simulation results in Figure S3 to compare our Model A with Models C and E, which does not use co-data at all and completely pools the data from trial $\mathcal{T}_1$ to $\mathcal{T}_2$. Numerical results of the complete comparison based on Models A – E are listed in Table S2.

In scenarios 1 – 3 when the bridging assumption is correct and/or animal data provide high predictability of human toxicity, it is evident that Model A leads to significantly increased PCS and allocates more patients to the true MTD in both regions $\mathcal{R}_1$ and $\mathcal{R}_2$. For example, comparing Models A and C, the PCS was increased from 14.7% to 42% in scenario 1 for trial $\mathcal{T}_1$ due to the use of animal data, and from 17% to 45% for trial $\mathcal{T}_2$ due to the use of animal data and correct bridging strategy. Model E does not leverage animal data into trial $\mathcal{T}_1$ but completely pools in the $\mathcal{T}_1$ trial data into trial $\mathcal{T}_2$. As a result, we observe that the PCS for trial $\mathcal{T}_1$ was much lower than that of Model A in these scenarios, although a higher PCS is observed in scenario 2 for pooling in the consitent $\mathcal{T}_1$ trial data. Using trial data from other human subgroup by such completely pooling approach could be harmful, which is evidenced by the results in scenario 6. Comparing Model A and Model E, we see that the latter allocated much more patients to excessively toxic doses which are 5, 10 and 20 mg/kg in this scenario, and more often decleared dose 1 mg/kg as the MTD. This is not surprising: pooling in the $\mathcal{T}_1$ trial data means we suppose patients in regions $\mathcal{R}_1$ and $\mathcal{R}_2$ are exchangeable, and Model E would consequently underestimate the human toxicity in trial $\mathcal{T}_2$.

We now switch our focus to assessing the accuracy of the posterior point estimates based on the Bayesian analysis models. As has been illustrated, leveraging co-data that are (in)consistent with current trial data, should improve the accuracy of posterior estimates for the probability of

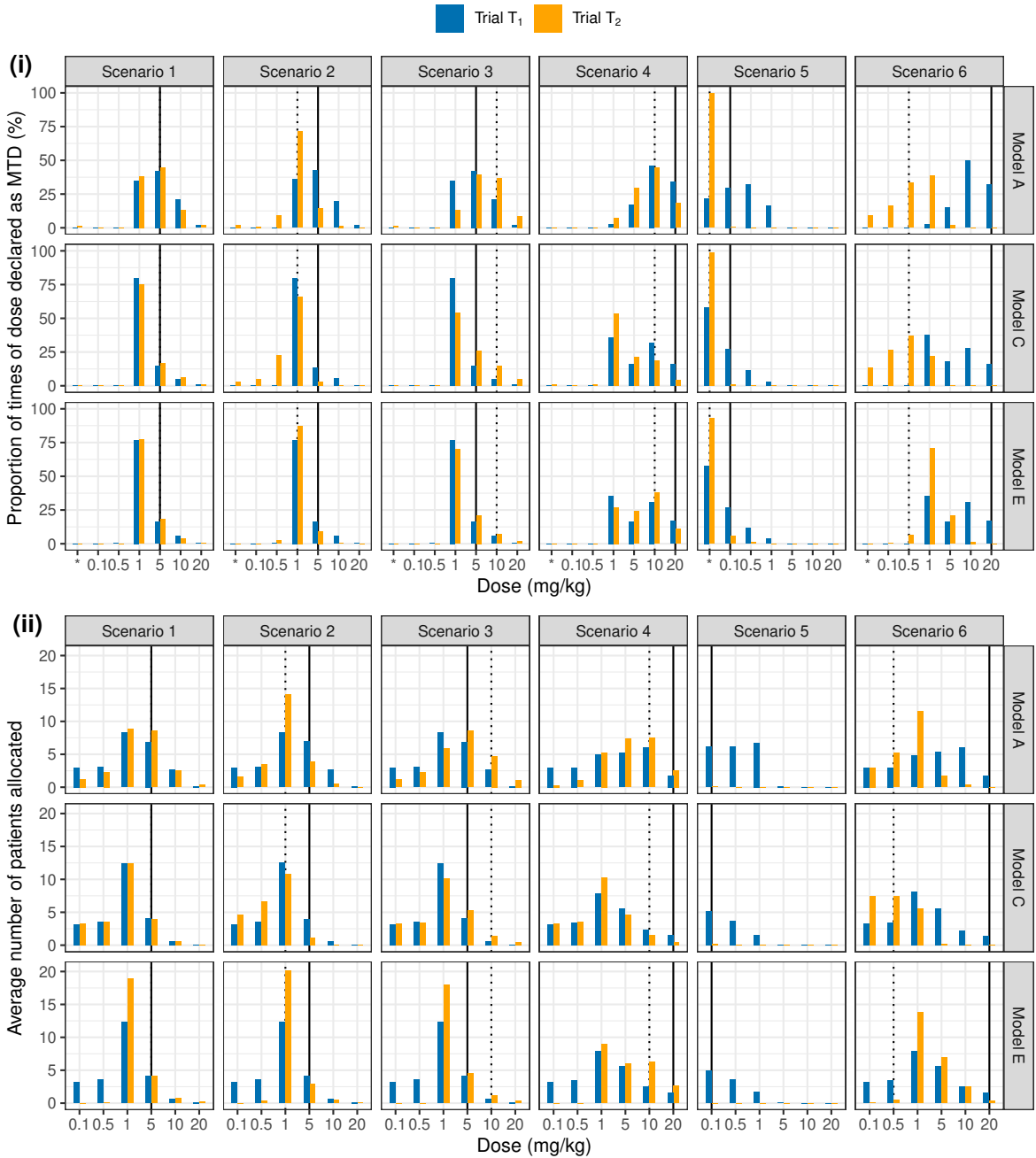

Figure S3: Operating characteristics of the adaptive phase I dose-escalation trials in two geographic regions, conducted and analysed using Models A, C, D. The vertical black solid (dotted) line indicates the true MTD in the western trial $\ell^\star$ (eastern trial $\ell$) under each simulation scenario.

toxicity. Analysis models that enable borrowing of information in this situation correspondingly will present (dis)advantages over those not permitting borrowing at all. We are thus much concerned with the accuracy of posterior estimates based on the proposed Bayesian model compared with its alternatives. In the simulation study, we preserve the point estimates (posterior medians) by the end of the *completed* trials, which may be a subset of or the entire 1000 pairs. We average across such point estimates per human dose under the analysis models A – E to approximate the


\end{document}